\documentclass[prb,twocolumn,showkeys,showpacs,amsmath,amsfonts,floatfix,superscriptaddress,nofootinbib,longbibliography]{revtex4-2}

\usepackage{amssymb,amsmath,amstext}                
\usepackage{graphicx}                                               
\usepackage{epstopdf}                                               
\usepackage{color}       
\usepackage{bm}
\usepackage{appendix}                                              
\usepackage[latin2]{inputenc}
\usepackage{ulem}
\usepackage{latexsym}
\usepackage[colorlinks=true,citecolor=blue,linkcolor=magenta]{hyperref}
\def\be{\begin{equation}}
\def\ee{\end{equation}}
\def\bea{\begin{eqnarray}}
\def\eea{\end{eqnarray}}
\def\bi{\begin{itemize}}
\def\ei{\end{itemize}}

\newcommand{\bra}[1]{\mbox{$\langle #1 |$}}
\newcommand{\ket}[1]{\mbox{$| #1 \rangle$}}

\begin{document}

\title{  Finite temperature tensor network study of \\
         the Hubbard model on an infinite square lattice }

\author{Aritra Sinha} 
\affiliation{Jagiellonian University, Institute of Theoretical Physics, 
             ulica \L{}ojasiewicza 11, 30-348 Krak\'ow, Poland }  
             
\author{Marek M. Rams} 
\affiliation{Jagiellonian University, Institute of Theoretical Physics, 
             ulica \L{}ojasiewicza 11, 30-348 Krak\'ow, Poland }       

\author{Piotr Czarnik} 
\affiliation{Jagiellonian University, Institute of Theoretical Physics, 
             ulica \L{}ojasiewicza 11, 30-348 Krak\'ow, Poland }
\affiliation{Theoretical Division, Los Alamos National Laboratory, Los Alamos, New Mexico 87545, USA.}
 
\author{Jacek Dziarmaga} 
\affiliation{Jagiellonian University, Institute of Theoretical Physics, 
             ulica \L{}ojasiewicza 11, 30-348 Krak\'ow, Poland }

\date{\today}

\begin{abstract}

The Hubbard model is a longstanding problem in the theory of strongly correlated electrons and a very active one in the experiments with ultracold fermionic atoms. 
Motivated by current and prospective quantum simulations, we apply a two-dimensional tensor network---an infinite projected entangled pair state---evolved in imaginary time by the neighborhood tensor update algorithm working directly in the thermodynamic limit.
With $U(1)\times U(1)$ symmetry and the bond dimensions up to 29, we generate thermal states down to the temperature of $0.17$ times the hopping rate.
We obtain results for spin and charge correlators, unaffected by boundary effects. The spin correlators---measurable in prospective ultracold atoms experiments attempting to approach the thermodynamic limit---provide evidence of disruption of the antiferromagnetic background with mobile holes in a slightly doped Hubbard model. 
The charge correlators reveal the presence of hole-doublon pairs near half filling and signatures of hole-hole repulsion on doping.
We also obtain specific heat in the slightly doped regime.
\end{abstract}

\maketitle

\section{The Hubbard model} 
One of the simplest models of interacting fermions on a lattice is the Fermi Hubbard model (FHM) with on-site repulsion between electrons of opposite spins: 
\bea
H &=& 
- \sum_{\langle i,j \rangle\sigma} 
  t\left( c_{i\sigma}^\dag c_{j\sigma} + c_{j\sigma}^\dag c_{i\sigma} \right) + \nonumber\\
  & &
  \sum_i U \left( n_{i\uparrow} - \frac12 \right) \left( n_{i\downarrow} - \frac12 \right) -
  \sum_i \mu \; n_i.
\label{H}
\eea
Here $c_{i\sigma}$ annihilates an electron with spin $\sigma=\uparrow,\downarrow$ at site $i$, $n_{i\sigma}=c^\dag_{i\sigma}c_{i\sigma}$ is the number operator, $n_i=n_{i\uparrow}+n_{i\downarrow}$, repulsion strength $U>0$, and $\mu$ is the chemical potential. Here $\langle i,j \rangle$ denotes summation over nearest-neighbor (NN) sites on a square lattice with hopping energy $t>0$. Although FHM is deemed to be an inordinately simple model for describing real materials, the competition between $t$ and $U$ gives rise to a myriad of physical phenomena, including stripe phases and Mott insulator. The model has exact solutions for  some limits in one dimension~\cite{Lieb1968, Lieb2003}. However, obtaining thermodynamic results for a two-dimensional (2D) system is exceedingly challenging, even with the most sophisticated numerical techniques, see~\cite{Qin2022} for a recent review. 

On the experimental front, ultracold atoms serve as a simulation platform where one can realize condensed matter physics models with high tunability~\cite{lewenstein2007ultracold, Bloch2008, Bloch2012}, including FHM~\cite{Strohmaier2007,
schneider2008,jordens2008mott, Jordens2010, Esslinger2010, Tarruel2018, Hofstetter2018, bohrdt2021exploration}. 
Quantum gas microscopy~\cite{Bakr2009, Sherson2010, Stefan2016, Gross2021} promises manipulations of individual atoms in optical lattices with faithful spin and density readouts, and have achieved impressive success in simulation of the many-body physics of fermions with alkali, potassium, and lithium isotopes~\cite{Cheuk2015, Haller2015, Parsons2015, Edge2015}. Soon followed single-site resolved detection of 2D Fermi-Hubbard physics encompassing imaging of antiferromagnetic correlations~\cite{Greif2016, Cheuk2016, Parsons2016, Boll2016, Brown2017}, entanglement entropy~\cite{Islam2015, Kaufman2016, Lukin2019, Rispoli2019}, hidden string order and magnetic polarons~\cite{Timon2017, Koepsell2019, Salomon2019, chiu19}.  These experiments mostly use harmonically confined systems of tens of fermions and can reach temperatures down to $T/t=0.25$. 

\begin{figure}[t!]
\vspace{-0cm}
\includegraphics[width=0.999\columnwidth,clip=true]{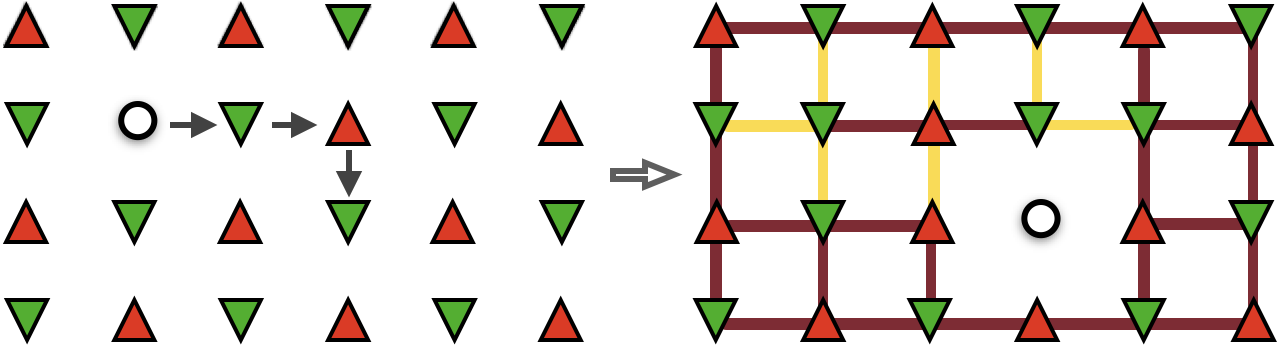}
\vspace{-0cm}
\caption{
{\bf Hole motion in AFM background. }
Illustration of an anti-ferromagnetic 2D lattice of spins with an alternating arrangement of spin-$\uparrow$ (red triangle) and spin-$\downarrow$ (green triangle). 
In the left diagram, just below the top left corner of the lattice, a hole is placed. We indicate its possible trajectory with arrows. 
In the right diagram, the hole is displaced by three lattice sites. The brown (yellow) lines indicate NN anti-ferromagnetic (ferromagnetic) correlations, capturing the disruption of the background AFM order by the hole motion.}
\label{fig:sketch2}
\end{figure}
In FHM with low doping, a hole moving in an anti-ferromagnetic (AFM) background leaves a track of ferromagnetically bound bubbles known as magnetic polarons, see Fig.~\ref{fig:sketch2}. Direct experimental detection of magnetic polarons~\cite{Koepsell2019} has fuelled several recent experimental and numerical/theoretical efforts~\cite{Grusdt2018, Grusdt2019, Blomquist2020, Bohrdt2020,  Ji2021, Bohrdt2021,Nielsen2021, Wrzosek2021, Lo2022, Kale2022}. The latter frequently resort to the effective $t{-}J$ model~\cite{Chao_1977}, or study a rather particular single-hole doping limit. Our work provides results for finite temperature spin and charge correlations for the 2D FHM directly in the thermodynamic limit. It is in line with recent experimental efforts to push quantum simulation of FHM towards the same limit by trapping hundreds of ultracold atoms in a box-like potential~\cite{chiu19}.

Here we consider $U/t=8$ and an average electron density per site $n= \langle  n_i\rangle=0.875$ and $1$, which correspond to doping $p=1-n=0.125$ and $0$, respectively. For these values, FHM captures essential aspects of high-$T_c$ superconductors such as the stripe phases~\cite{Simons_Hubb_17, stripes_finite_mets,stripes_AFQMC}, although additional terms such as next NN hopping might be necessary to stabilize the superconducting phase~\cite{Boris2019} and further additional bands for fluctuating stripes at finite temperature~\cite{huang17}. In the following, we set $t=1$ as a unit of energy. The temperature is measured in these units ($k_B = 1$).

\section{Tensor networks} 
Quantum condensed matter states hosted by two-dimensional lattices can often be efficiently represented by a type of tensor networks (TN)~\cite{Verstraete_review_08,Orus_review_14} known as the infinite projected entangled-pairs state (iPEPS) ansatz~\cite{Nishino_2DvarTN_04,Verstraete_PEPS_04, Murg_finitePEPS_07}. It is a state-of-the art numerical method for strongly correlated systems~\cite{Cirac_iPEPS_08,Xiang_SU_08,Gu_TERG_08,Orus_CTM_09}. 
The iPEPS was instrumental in solving the long-standing magnetization plateaus problem in the highly frustrated compound $\textrm{SrCu}_2(\textrm{BO}_3)_2$~\cite{matsuda13,corboz14_shastry}, establishing the striped nature of the ground state of the doped 2D Hubbard model~\cite{Simons_Hubb_17}, and providing new evidence for gapless spin liquid in the kagome Heisenberg antiferromagnet~\cite{Xinag_kagome_17,TRG_kagome_Wen}. Further technical progress~\cite{fu,Corboz_varopt_16,Vanderstraeten_varopt_16,Fishman_FPCTM_17,Xie_PEPScontr_17,Corboz_Eextrap_16,Corboz_FCLS_18,Rader_FCLS_18,Rams_xiD_18} paved the way towards even more challenging problems, including simulation of thermal states~\cite{Czarnik_evproj_12,Czarnik_fevproj_14,Czarnik_SCevproj_15, Czarnik_compass_16,Czarnik_VTNR_15,Czarnik_fVTNR_16,Czarnik_eg_17,Dai_fidelity_17,CzarnikDziarmagaCorboz,czarnik19b,Orus_SUfiniteT_18,CzarnikKH,jimenez20,CzarnikSS,Poilblanc_thermal}. Very recently there has been promising advancements to calculate the ground states of three-dimensional systems~\cite{Vlaar2021}.
The alternative tensor-network-based approach considers systems on cylinders and is now routinely used to investigate 2D ground states using density matrix renormalization group (DMRG) ~\cite{Simons_Hubb_17,CincioVidal} and was also applied to thermal states~\cite{Stoudenmire_2DMETTS_17,Weichselbaum_Tdec_18, WeichselbaumTriangular,WeichselbaumBenchmark,Wei2021, Abanin2016,stripes_finite_mets}. It is, nevertheless, severely limited by the exponential growth of the bond dimension with system's width. Furthermore, tensor network  approaches  relying on  contraction of a 3D tensor network  representing  a 2D  thermal state have been proposed~\cite{Li_LTRG_11,Xie_HOSRG_12,Ran_ODTNS_12,Ran_NCD_13,Ran_THAFstar_18,Su_THAFoctakagome_17,Su_THAFkagome_17,Ran_Tembedding_18}.

\begin{figure}[t!]
\vspace{-0cm}
\includegraphics[width=0.999\columnwidth,clip=true]{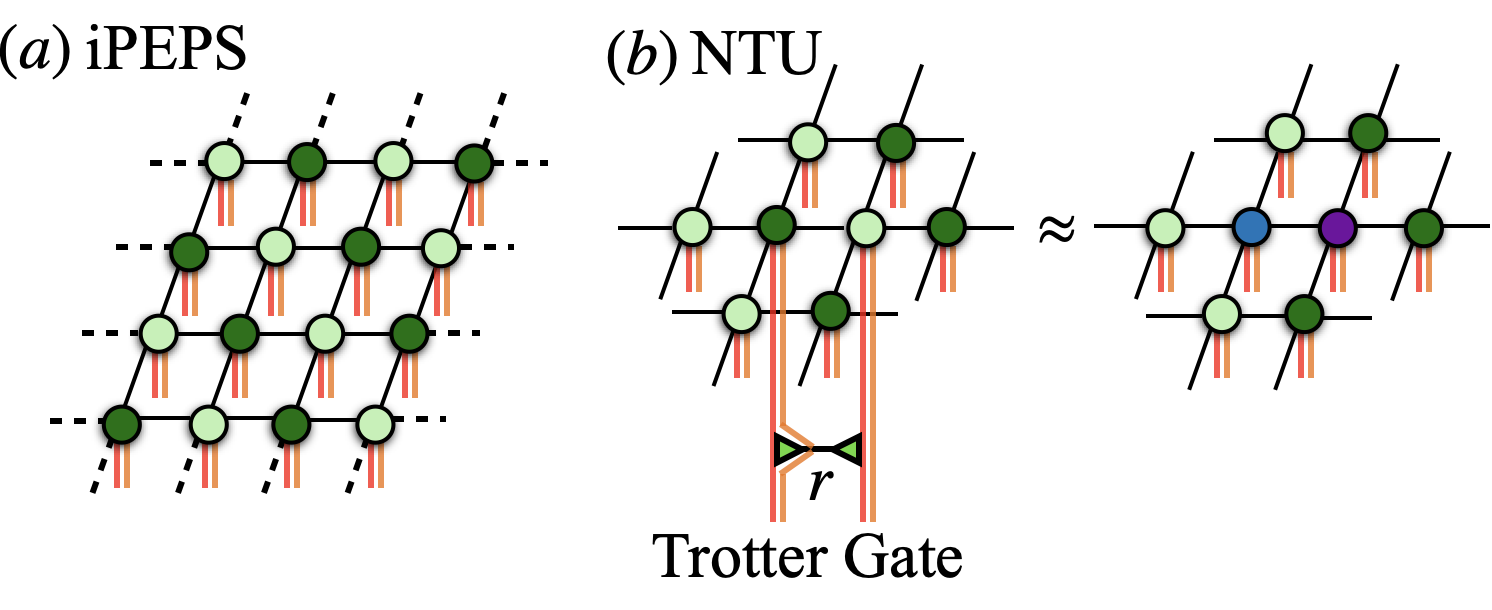}
\vspace{-0cm}
\caption{
{\bf iPEPS ansatz and the neighborhood tensor update. }
(a)Representation of an infinite PEPS tensor network with tensors $A$ (lighter green) and $B$ (darker green) on the checkerboard lattice. It represents the purification of a thermal state where the red and orange lines indicate physical and ancillary indices respectively, and the black lines are bond indices of bond dimension D connecting NN sites.
(b)Application of a Trotter gate to a horizontal NN pair of $A$-$B$ tensors. The gate can be viewed as a contraction of two tensors by an index with dimension $r$. When those two tensors get absorbed into tensors $A$ and $B$, the bond dimension connecting them increases to $r\times D$. On the right hand site, they are approximated by a pair of new tensors, $A'$ (blue) and $B'$ (purple), having the original bond dimension $D$. In NTU, the new tensors are optimized to minimize the difference between the presented clusters. Subsequently, they form a new checkerboard network after the update.  
\label{fig:sketch1}}
\end{figure}

In this paper, we apply iPEPS with abelian symmetries(see Appendix~\ref{app:iPEPS} for brief description) to the 2D Hubbard model at finite temperature. We perform direct imaginary time evolution of an iPEPS that represents purification of the thermal state~\cite{CzarnikDziarmagaCorboz}, see Fig.~\ref{fig:sketch1}(a). For the sake of its numerical stability, we use a fermionic version of the neighborhood tensor update (NTU) algorithm~\cite{ntu}, implementing the $U(1)\times U(1)$ symmetry and further refinements. We enforce fermionic statistics by following a general scheme of Refs.~\cite{Corboz09_fmera,Corboz_fiPEPS_10}.
The latter include the spatially and rotationally invariant assignment of symmetry sectors' bond dimensions of the tensors (FIX) and an environment-assisted truncation (EAT) procedure which makes the Trotter step of the NTU algorithm better aware of its tensor environment. We provide detailed descriptions of FIX and EAT in Appendix~\ref{app:EAT}. 
We calculate two-point spin and charge correlators for inverse temperatures $0<\beta<6$. These results are well converged in the iPEPS bond dimension and, by construction, they are free of finite-size effects. This range of temperatures is accessible for current ultracold atoms experiments attempting to reach the thermodynamic limit. 

The imaginary time evolution of the purification, $e^{-\frac12\beta H}$, is performed by the second-order Suzuki-Trotter decomposition of small time steps. An application of the NN two-site Trotter gate is outlined in Fig.~\ref{fig:sketch1}(b) and in Appendix~\ref{app:NTU}. After a Trotter gate is applied to a bond connecting NN sites, its bond dimension is increased by a factor equal to the singular value decomposition (SVD) rank of the gate, $r$. In order to prevent its exponential growth, the dimension is truncated with NTU to its original value, $D$, in a way that minimizes the truncation error. NTU~\cite{ntu,mbl_ntu} can be regarded as a special case of a cluster update~\cite{wang2011cluster,Lubasch_cluster_1,Lubasch_cluster_2}, where the number of neighboring lattice sites taken into account during truncation makes for a refining parameter. The cluster update interpolates between a local truncation---as in the simple update (SU)---and the full update (FU) that takes into account all correlations in the truncated state~\cite{CzarnikDziarmagaCorboz}. 
As the NTU cluster includes the neighboring sites only, see Fig.~\ref{fig:sketch1}(b), the NTU error can be calculated numerically exactly via parallelizable tensor contractions~\cite{ntu,mbl_ntu}. We provide short description of the algorithm in Appendix~\ref{app:NTU}. That exactness warrants that the error measure is Hermitian and
non-negative down to the numerical precision, unlike in the case of FU that involves the approximate corner transfer matrix renormalization group (CTMRG)~\cite{Baxter_CTM_78, nishino1996, Orus_CTM_09, Corboz_CTM_14}. It is thus an optimal trade-off for applications where quantum correlations are not too long like in, e.g.,
Kibble-Zurek quenches in 2D~\cite{KZ2D} or time evolution of many-body localizing systems~\cite{mbl_ntu}. 
Therefore, it should perform well for the Hubbard model at intermediate and high temperatures, as we demonstrate in Appendix~\ref{app:benchmarks} using spinless non-interacting fermions and available DCA results for FHM~\cite{leblanc15}.
We apply iPEPS with bond dimensions up to $29$ to the Hubbard model directly in the thermodynamic limit in a regime complementary to iPEPS simulations at zero temperature~\cite{Corboz_stripes_11,Simons_Hubb_17}, exponential thermal renormalization group (XTRG) of a small square lattice~\cite{Wei2021}, or minimally entangled typical thermal states (METTS) on thin cylinders~\cite{stripes_finite_mets}.

\begin{figure}[t!]
\vspace{-0cm}
\includegraphics[width=\columnwidth,clip=true]{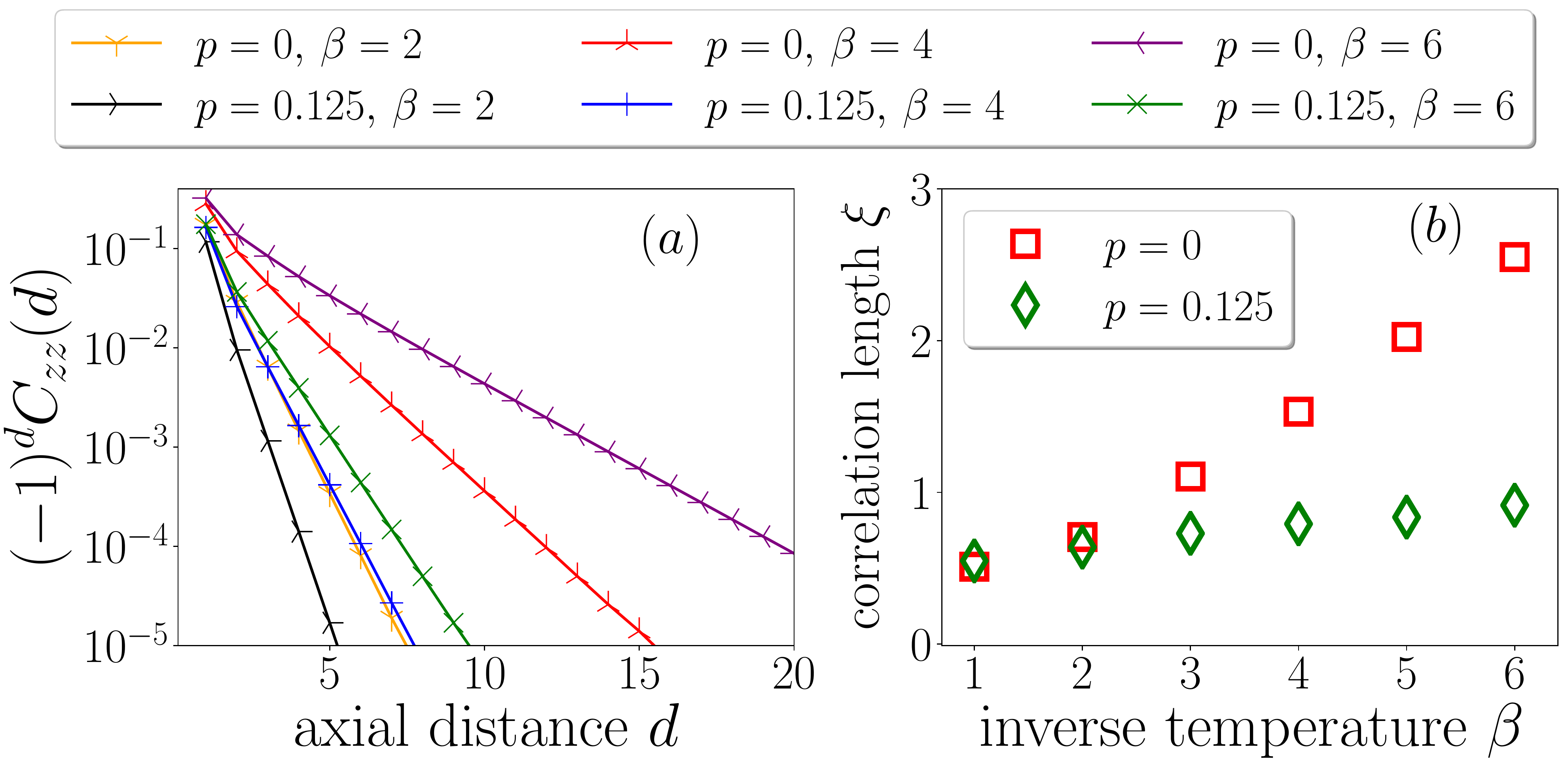}
\vspace{-0cm}
\caption{
{\bf Long-range correlators and the correlation length.} 
In (a), long range staggered spin-spin correlators along $z$-direction,
$(-1)^d C_{zz}(d)$, are plotted for three values of the inverse temperature: $\beta=2, 4$ and $6$. Finite values of $(-1)^d C_{zz}(d)$ for $d\gg 1$ shows the presence of
anti-ferromagnetic correlations across a finite range. In (b), we plot the correlation length $\xi$, characterizing $C_{zz}(d)$, versus $\beta$. }
\label{fig:long_correlator}
\end{figure}

\section{Results} 
At half-filling and for large on-site Coulomb repulsion $U\gg1$, FHM can be mapped to the Heisenberg model. The Heisenberg model develops long-range AFM order at zero temperature and strong correlations persist even at moderate temperatures~\cite{Manou1991}. In case of FHM, these AFM correlations do not perish even at low doping for low and intermediate temperatures~\cite{mazurenko17}. We corroborate these findings with our iPEPS simulations by calculating two-site spin-spin correlation function along the $z$-direction: $C_{zz}(d) = \left< Z_{i}Z_{i+d}\right>- \left<Z_{i}\right> \left<Z_{i+d}\right>$, where $Z_{i} =  n_{i\uparrow} - n_{i\downarrow}$ and $d$ is the distance along axial direction. In Fig.~\ref{fig:long_correlator}(a), we plot the correlators for half-filling ($p=0$), and $p=0.125$ at $\beta=2, 4, 6$. 

In Fig.~\ref{fig:long_correlator}(b), we plot the $\beta$-dependence of the axial correlation length $\xi$, characterizing $C_{zz}(d)$, extracted from a transfer matrix (see Appendix~\ref{app:benchmarks} for details). $\xi\approx2.5$ that we reach for $p=0$ would be problematic on a thin cylinder or in a small system. The length is shorter for doping $p=0.125$ where doping undermines the AFM order. We verified the convergence of the results with the bond dimension of PEPS tensors.

\begin{figure}[t]
\vspace{-0cm}
\includegraphics[width=\columnwidth,clip=true]{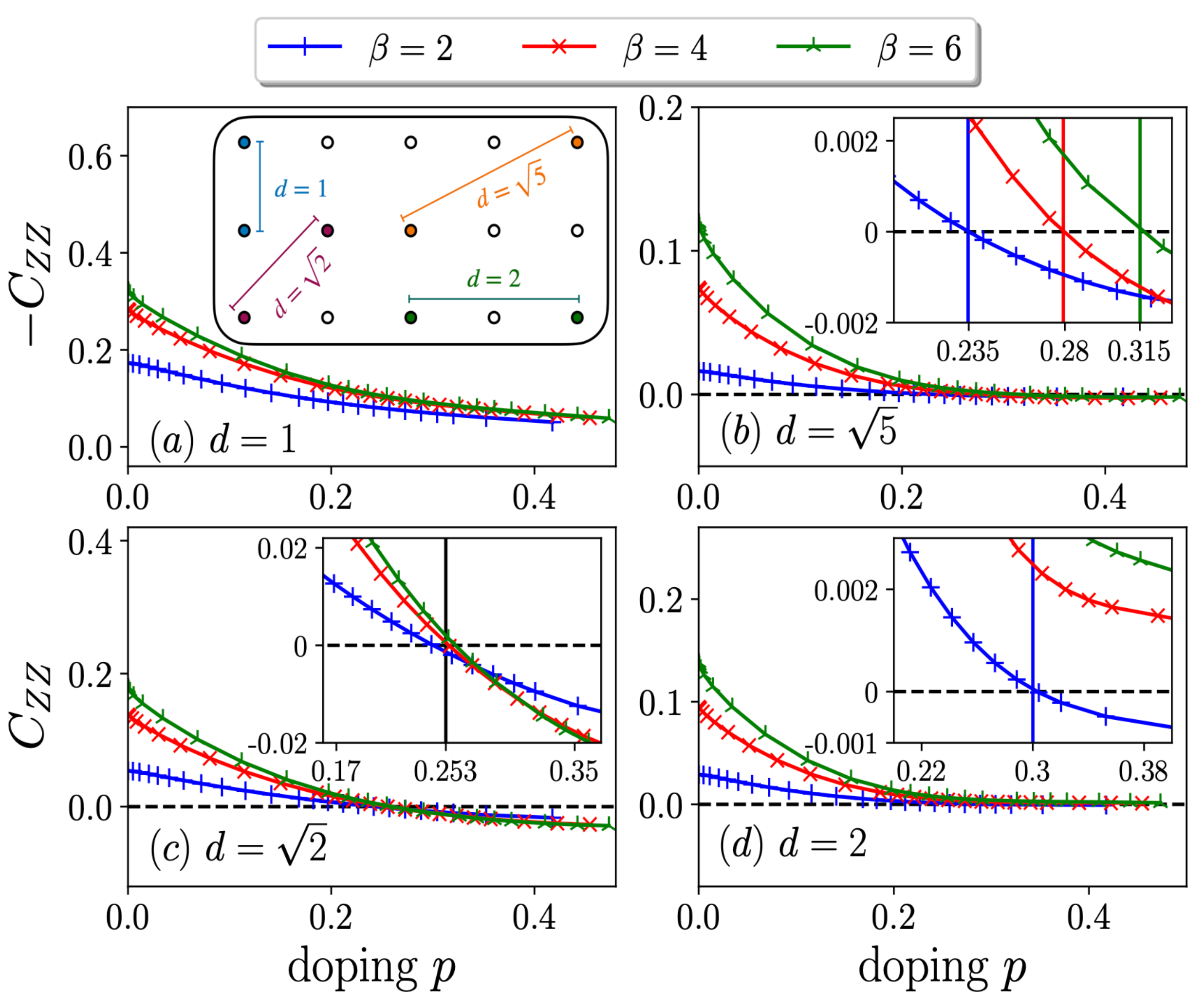}
\vspace{-0cm}

\caption{{\bf Spin correlators for varying doping.} 
We show the staggered (a) nearest axial neighbor($d=1$),  (b) next nearest diagonal ($d=\sqrt{5}$), (c) next nearest axial neighbor ($d=2$), and (d) nearest diagonal ($d=\sqrt{2}$) spin-spin correlators along the $z$-direction $(-1)^{l(d)} C_{zz}(d)$ for $U=8$ as a function of doping $p$ and for three values of the inverse temperatures: $\beta=2,4,6$. 
Here $l(d)$ is the Manhattan distance between lattice sites, see the inset of (a) for visual aid. The positive values of the sign-corrected correlators are consistent with their antiferromagnetic ordering. The insets of (b), (c), and (d) reveal the doping around which the correlators change signs. The latter signals decay in antiferromagnetic order. }
\label{fig:short_spin_correlator}
\end{figure}

In Fig.~\ref{fig:short_spin_correlator} we present several short-range correlators: $C_{zz}(d=1)$ (nearest axial correlator), $C_{zz}(d=\sqrt{2})$ (nearest diagonal correlator),  $C_{zz}(d=2)$ (next nearest axial correlator), and $C_{zz}(d=\sqrt{5})$ (next nearest diagonal correlator) in the function of a doping. It is motivated by a recent experiment~\cite{chiu19} in a small system of ultra-cold atoms, where a change of sign in $C_{zz}(d=\sqrt{2})$ has been observed around doping $p=0.2$, that is not inconsistent with numerical XTRG study of small lattices up to $8\times 8$ sites~\cite{Wei2021}. Our thermodynamic limit results in Fig.~\ref{fig:short_spin_correlator} further validate this effect. We also observe monotonic decreasing of the correlator with doping with no characteristic minima seen in the finite-system data.
It has been long noted that a hole traveling through a strongly-coupled Hubbard model near the half-filling leaves behind a trail of ferromagnetically ordered regions in the AFM background~\cite{nagoka1966, nagaev1968, Brinkman1970, Trugman1988}, as illustrated in Fig.~\ref{fig:sketch2}. 
One expects that the magnitude of two-point correlators decreases with increasing doping since the interaction between holes and the AFM background creates magnetic polarons. This phenomenon is captured particularly well through the sign reversal in diagonal correlators and has been argued~\cite{chiu19} to be explained by the geometric string theory~\cite{Grusdt2018}. We recover this feature in our simulations in Fig.~\ref{fig:short_spin_correlator}(b), where we observe a change of sign around $p=0.253\pm 0.007$ for all three inverse temperatures, $\beta=2, 4, 6$. 

Other correlators, $C_{zz}(d=1)$ and $C_{zz}(d=2)$, show qualitative similarity with those obtained in~\cite{Wei2021}, albeit without suffering from finite-size effects. Furthermore, we obtain longer-range correlators like, e.g., the second diagonal correlator $C_{zz}(d=\sqrt{5})$. It undergoes a similar change of sign at $0.25<p<0.32$ for all three temperatures providing another strong validation of the string theory~\cite{chiu19}. We obtain converged values of doping $p$ (where the sign changes) by working in the thermodynamic limit with iPEPS. It could potentially help benchmark other numerical methods and experiments. Note that due to finite size effects, previous approaches have been unable to obtain sharp estimates for this sign reversal.

\begin{figure}[t]
\vspace{-0cm}
\includegraphics[width=\columnwidth,clip=true]{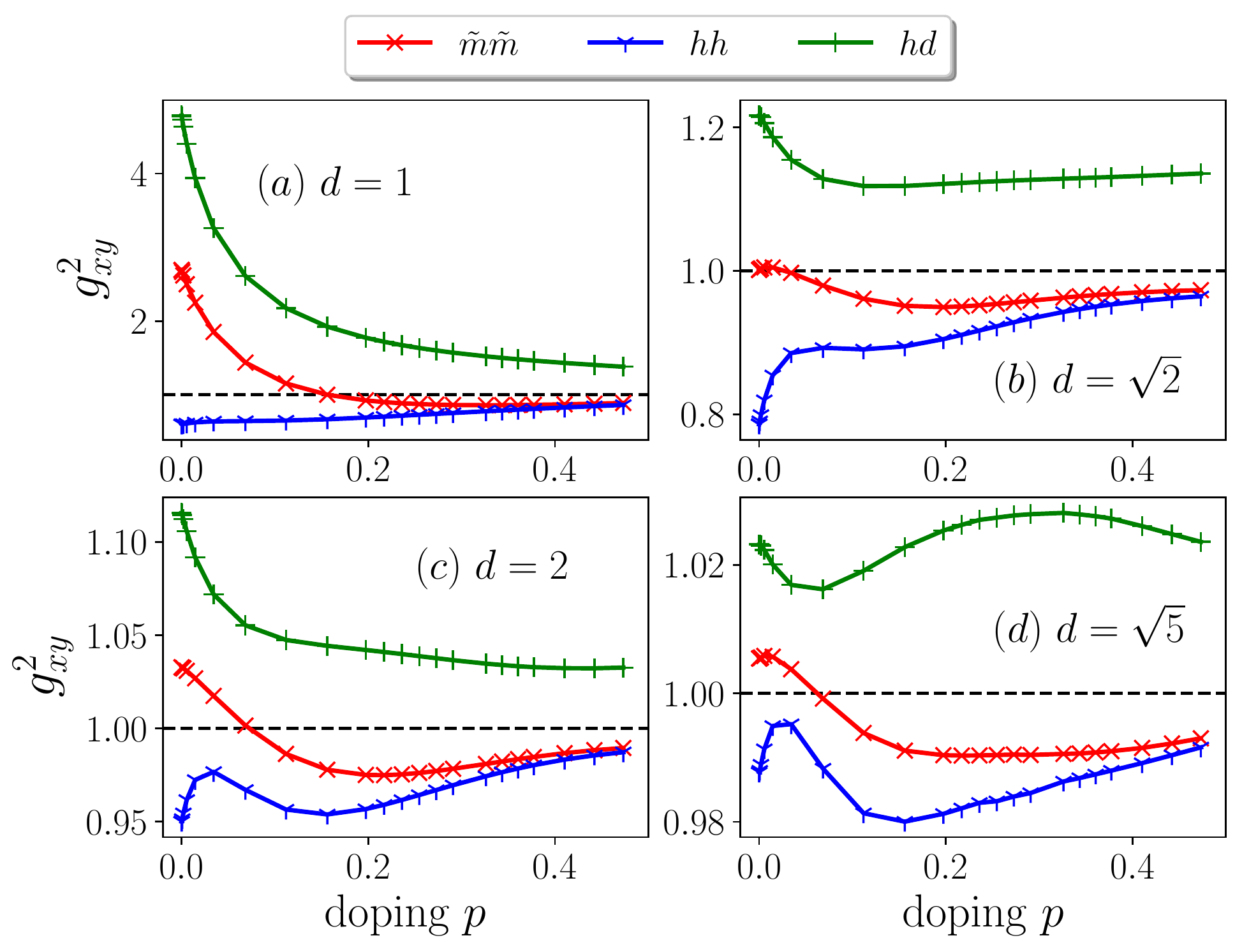}
\vspace{-0cm}
\caption{{\bf Normalized charge correlators for varying doping.} 
Normalized hole-hole, hole-doublon and anti-moment correlators for inverse temperature $\beta=6$ are plotted versus doping $p$. The meaning of $xy$ in $g_{xy}$ is given in the legend. Normalized anti-moment correlators, $g_{2}^{\tilde{m}\tilde{m}}$, show bunching at small doping and anti-bunching at large dopings, while normalized hole-doublon, $g_{2}^{hd}$, and hole-hole correlators, $g_{2}^{hh}$, always show bunching and anti-bunching respectively.}
\label{fig:g2_short_correlator}
\end{figure}

Important characteristics of the doped FHM can also be revealed by charge correlators. The interesting two-point charge correlators usually  considered are the hole-hole, $h_{i}h_{i+d}$, correlator  where hole $h_{i} = (1- n_{i\uparrow})(1-n_{i\downarrow})$  and hole-doublon correlator, $h_{i}d_{i+d}$, where $d_{i} = n_{i\uparrow}n_{i\downarrow}$.
The quantum gas microscopy techniques overestimate and cannot distinguish between holes and doublons, as they both appear the same after imaging, but it can instead measure anti-moment correlators, $\tilde{m}_{i}\tilde{m}_{i+d}$, of $\tilde{m} = h+d$. Nevertheless, recent developments promise hole-doublon correlators measurements in the near-future~\cite{Hartke2020}. First, we calculate  normalized hole-hole $g^2_{hh}$, hole-doublon  $g^2_{hd}$, and anti-moment $g^2_{\tilde{m}\tilde{m}}$  correlation functions at inverse temperature $\beta=6$: 
\begin{equation}
g^2_{xy}(d) = 
\frac{\langle x_i  y_{i+d} \rangle}
{\langle x_i \rangle  \langle y_{i+d} \rangle },
\end{equation}
plotting the results in Fig.~\ref{fig:g2_short_correlator}. We find that both $g^{2}_{hd}(d=1)$ and $g^{2}_{\tilde{m}\tilde{m}}(d=1)$ show strong bunching near half filling ($p\to0$), indicating the presence of nearest neighbor hole-doublon pairs. This is further supported by the fact that beyond $d=1$, both  $g^{2}_{hd}$ and $g^{2}_{mm}$ show much weaker bunching effect. At high doping,  anti-bunching effects from hole-hole correlators $g^{2}_{hh}$ dominate and contribute to the cross-over of $g^{2}_{\tilde{m}\tilde{m}}$ correlators from bunching to anti-bunching. The behaviour of the correlators at $d=\sqrt{2}$, $2$ and $\sqrt{5}$ remains qualitatively similar to the ones at $d=1$, though much less pronounced. 
Our results are qualitatively consistent with finite-size experiments~\cite{Grusdt2018, Cheuk2016, chiu19, Hartke2020} and numerics~\cite{Wei2021}.

\begin{figure}[t]
\vspace{-0cm}
\includegraphics[width=\columnwidth,clip=true]{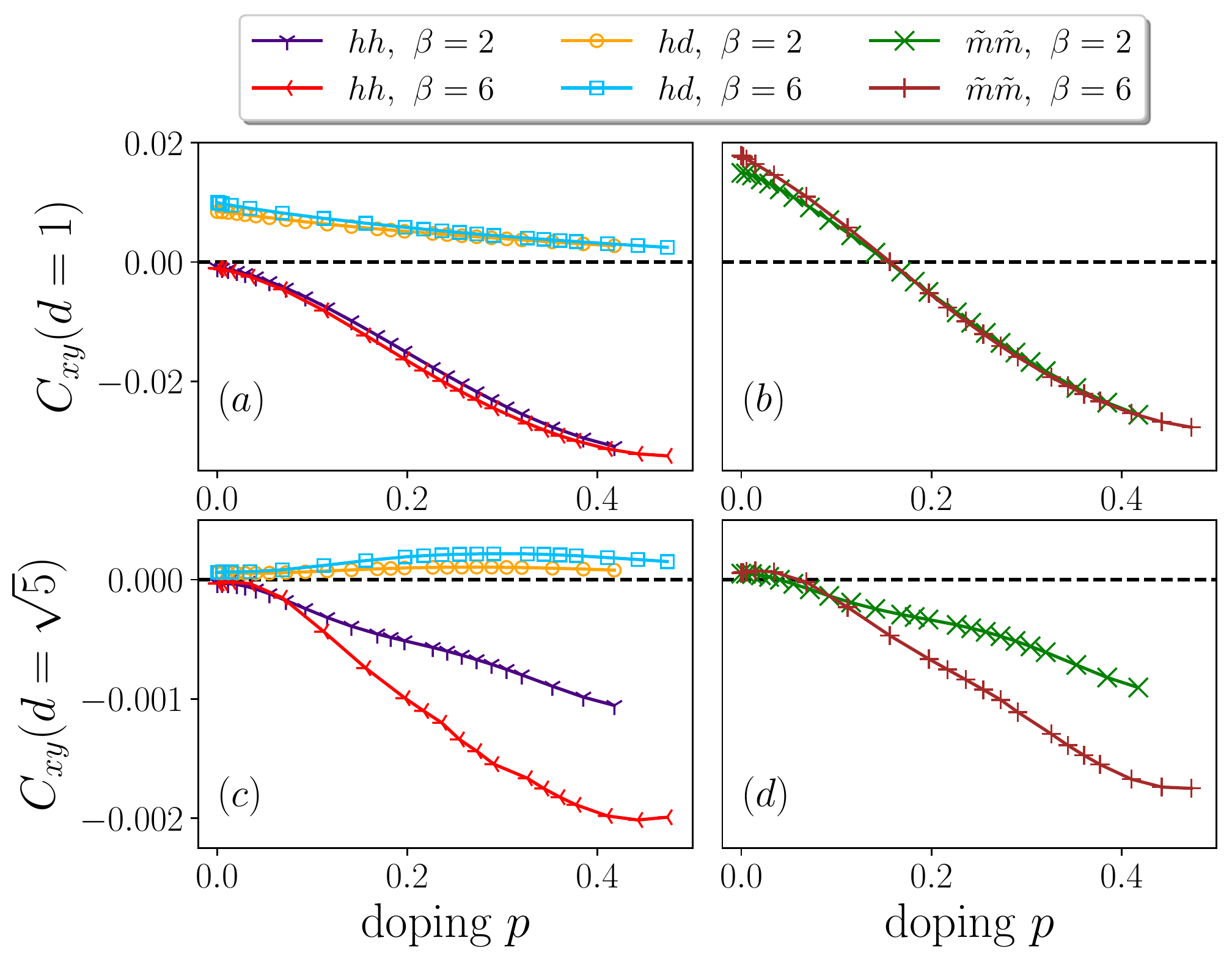}
\vspace{-0cm}
\caption{{\bf Connected charge correlators for varying doping.} 
We show connected hole-hole, $C_{hh}$, hole-doublon, $C_{hd}$, and anti-moment, $C_{\tilde{m}\tilde{m}}$, correlators for inverse temperatures $\beta=2$ and $6$ versus doping $p$. In (a, b) the results for NN ($d=1$) and in (c, d) next nearest diagonal ($d=\sqrt{5}$) correlators.  $C_{hh}(d=\sqrt{5})$ in (c) and $C_{\tilde{m}\tilde{m}}(d=\sqrt{5})$ in (d) show strong  temperature dependence.}
\label{fig:charge_correlator}
\end{figure}

Next, in Fig.~\ref{fig:charge_correlator}, we show connected hole-hole $C_{hh}$, hole-doublon $C_{hd}$ and anti-moment $C_{\tilde{m}\tilde{m}}$ correlation functions for axial nearest neighbor ($d=1$) and next nearest diagonal ($d=\sqrt{5}$):
\begin{equation}
C_{xy}(d) = {\langle x_i  y_{i+d} \rangle - \langle x_i \rangle  \langle y_{i+d} \rangle. }
\end{equation} We do not plot the doublon-doublon correlators $C_{dd}$ as their magnitude is relatively small, ${\cal O}(10^{-3})$, for NN correlators. It is important to note that, as $C_{\tilde{m}\tilde{m}} = C_{hh}+2C_{hd}+C_{dd}$, it is primarily the competition between $C_{hh}$ and $C_{hd}$ that drives the magnitude of $C_{\tilde{m}\tilde{m}}$. We see in Fig.~\ref{fig:charge_correlator} that as the system is gradually doped away from half-filling, the hole-doublon correlations decrease in magnitude while the hole-hole correlations increase significantly. Interestingly, $C_{hh}$ shows strong dependence on temperature beyond $d>1$, see Fig.~\ref{fig:charge_correlator}(c). For additional data on correlators, see the Appendix~\ref{app:corr}.

Finally, in Fig.~\ref{fig:cv}, we show specific heat as a function of the inverse temperature. We tune different chemical potentials to achieve desired doping of $p=0.125$ at each temperature point. We develop a method to control particle density by interpolating the chemical potential during imaginary-time simulation, see App.~\ref{app:interpolation} for details. In practice, however, we did not end up using it, as NTU evolution with different chemical potentials, avoiding computationally expensive CTMRG, can be executed more efficiently. The energy per site used here reads
\be 
E = 
-\frac12\sum_{j}\left(\left< c_{i\sigma}^\dag c_{j\sigma}\right> + \left< c_{j\sigma}^\dag c_{i\sigma}\right> \right) + 
U \left<n_{i\uparrow} n_{i\downarrow}\right> ,
\label{E}
\ee 
where $j$ runs over $4$ NN sites of the site $i$. Motivated by high-temperature expansion, numerical data for $E(\beta)$ were fitted with a polynomial in $\beta$ to stabilise numerical derivative. With increasing degree of the polynomial, but still far from overfitting, the specific heat converges to a curve with two peaks. For $p=0.125$, there is a sharp peak at $\beta_{1}=0.437$ and a broad one around $\beta_2=2$. The former, known as the \textit{charge peak}, is related to the suppression of the double occupancy. The latter is located near a point, which in the case of half-filling, corresponds to the crossover from a spin-disordered paramagnet to a state with NN antiferromagnetic correlations. It is known as the \textit{spin peak}. As the doping undermines the AFM order this crossover is less pronounced at $p=0.125$. Our results are in qualitative agreement with the quantum Monte Carlo on a $6\times 6$ cluster~\cite{CV_Hubbard} but they are free from finite-size effects.

\begin{figure}[t]
\vspace{-0cm}
\includegraphics[width=\columnwidth,clip=true]{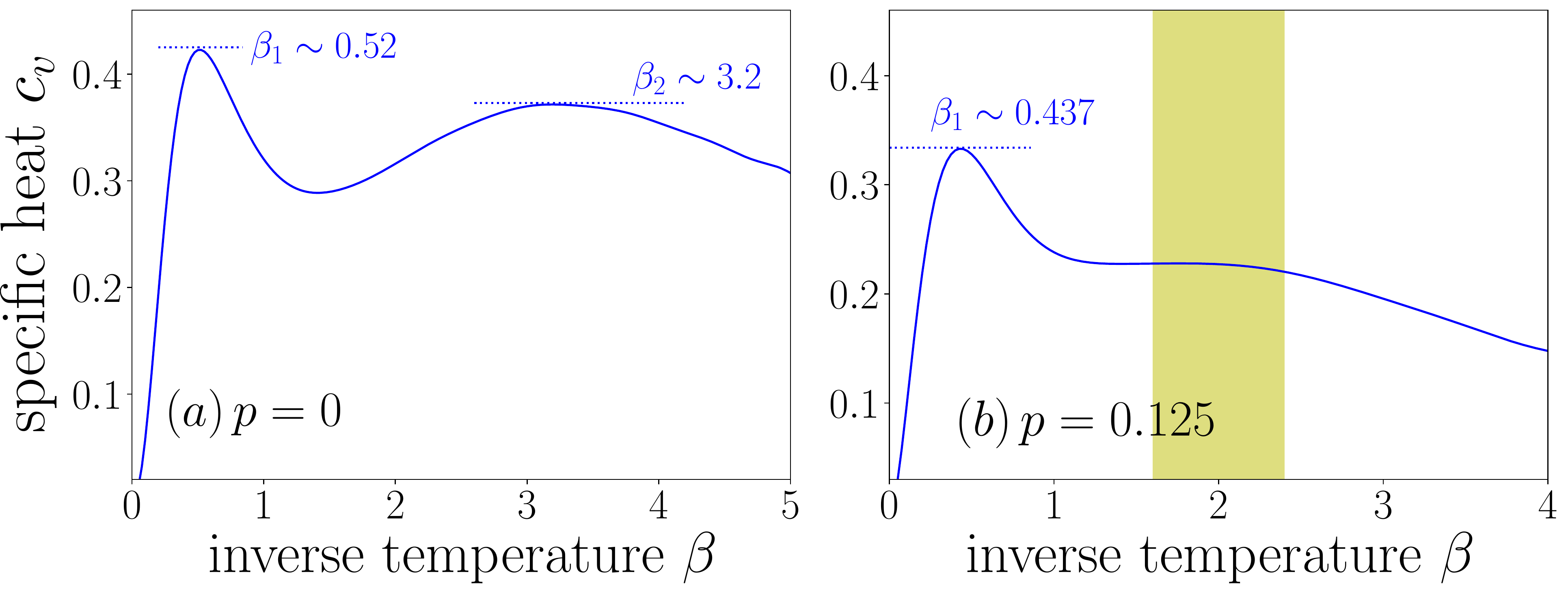}
\vspace{-0cm}
\caption{{\bf Specific heat.} 
The specific heat as a function of $\beta$ is obtained by a polynomial fit to numerical $E(\beta)$ for doping $p=0$ (a) and $p=0.125$ (b). The yellow shaded region marks the broad {\it spin peak} around $\beta=2$.}
\label{fig:cv}
\end{figure}

\section{Conclusion}
We extend the NTU algorithm to study spinful fermionic systems. We employ it to the challenging FHM in the 2D square infinite lattice at a finite temperature. We calculate expectation values for a set of observables that could be probed directly in prospective ultracold atoms experiments. By eliminating finite-size effects, we make contact with the current technology where large samples of atoms in almost homogeneous box-like trapping potentials can be probed. We cover a range of temperatures and dopings, including  those accessible to the current experiments.

\acknowledgements
%
AS is indebted to Gabriela Wojt\'{o}wicz, Titas Chanda and Juraj Hasik for useful discussions. We also thank Juraj Hasik and Krzysztof Wohlfeld for useful comments on the manuscript. PC acknowledges initial support from Laboratory Directed Research and Development (LDRD) program of Los Alamos National Laboratory (LANL) under project number 20190659PRD4 with subsequent support by by the National Science Centre (NCN), Poland under project 2019/35/B/ST3/01028. This research was supported in part by the National Science Centre (NCN), Poland under projects 2019/35/B/ST3/01028 (AS, JD) and 2020/38/E/ST3/00150 (MR).
%

\appendix
\renewcommand{\theequation}{\thesection\arabic{equation}}
\setcounter{equation}{0}
\renewcommand{\thefigure}{A\arabic{figure}}
\setcounter{figure}{0}
\renewcommand{\thetable}{A\arabic{table}}
\setcounter{table}{0}

\section{iPEPS and symmetries}
\label{app:iPEPS}


\begin{figure*}[t!]
\vspace{-0cm}
\includegraphics[width=0.5\textwidth,clip=true]{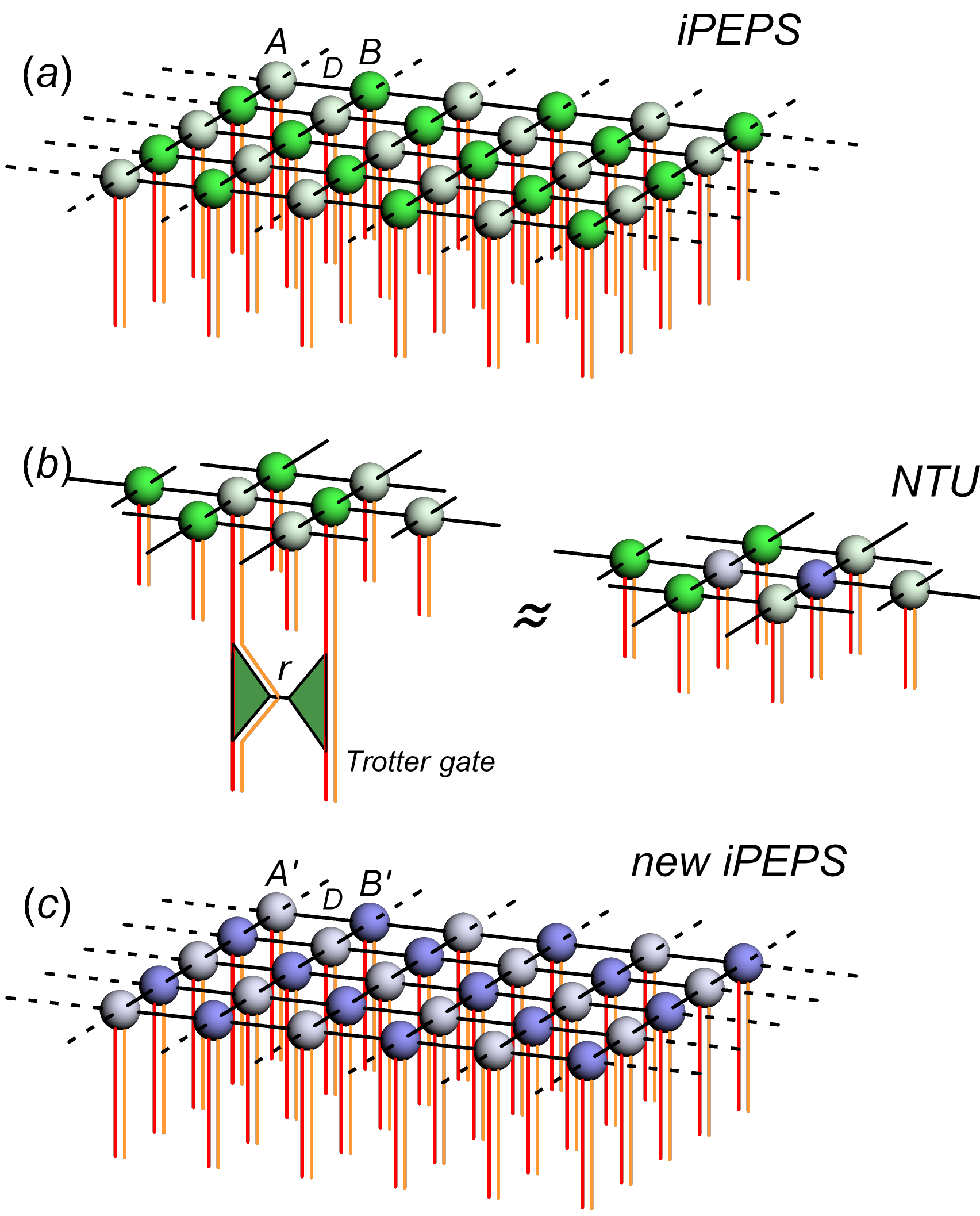}
\vspace{-0cm}
\caption{
{\bf Overview of NTU. }
In (a), we show infinite PEPS with tensors $A$ (lighter green) and $B$ (darker green) forming a checkerboard lattice. The iPEPS encodes a purification of a thermal state where the red/orange lines represent physical/ancilla indices, and the black lines are bond indices with total bond dimension $D$ connecting nearest neighbor sites.
In one of the Suzuki-Trotter steps, a Trotter gate is applied to physical indices of every horizontal nearest neighbor pair of $A$-$B$ tensors (but not to horizontal $B$-$A$ pairs). The gate is a contraction of two tensors by an index with dimension $r$. When the two tensors are absorbed into tensors $A$ and $B$, the bond dimension connecting them increases from $D$ to $r\cdot D$.
In panel (b), the $A$-$B$ pair -- with a Trotter gate applied to it -- is approximated by a pair of new tensors, $A'$ (lighter purple) and $B'$ (darker purple), connected by an index with the original dimension $D$. The new tensors are optimized to minimize the difference between the two presented tensor networks.
In (c),  $A'$ and $B'$ optimized in the last panel replace all tensors $A$ and $B$, forming a new iPEPS. 
Now, the next Trotter gate can be applied.
Each line crossing indicates the application of a SWAP gate, enforcing fermionic statistics.
}
\label{fig:NTU_overview}
\end{figure*}

The iPEPS ansatz used in this work assumes a checkerboard lattice of tensors with two sites, $A$ and $B$, in a unit cell. We depict it in Fig.~\ref{fig:NTU_overview}(a). Each iPEPS tensor has four legs containing virtual degrees of freedom---each with total bond dimension $D$, a physical index $s$, and an ancillary index $a$. The imaginary-time-evolved iPEPS $\ket{\psi(\beta)}$, which represents a purification of the thermal density operator $\rho(\beta)$, is obtained by an action of an evolution operator $U(\beta) = e^{-\frac{\beta}{2} H}$ on an uncorrelated product state at infinite temperature $\ket{\psi(\beta=0)}$. We choose the initial state $\ket{\psi(0)}$ to be a product of maximally entangled states of every physical site with its ancilla: $\ket{\psi(0)} = \prod_{j}\prod_{m=\uparrow,\downarrow} \frac{1}{\sqrt{2}} \sum_{s^m_j= a^m_j = 0,1} \ket{s^m_{j}a^m_{j}}$, where $j$ enumerates the lattice sites and $m$ refers to spin degrees of freedom---with two spin species at each lattice site for FHM. The density operator results from tracing out the ancillary degrees of freedom of the purification 

\bea 
\rho(\beta) \propto \exp(-\beta H) &=& \text{Tr}_{a}\ket{\psi(\beta)}\bra{\psi(\beta)}  \\
&=& \text{Tr}_{a}U(\beta)\ket{\psi(0)}\bra{\psi(0)}U(\beta).\nonumber
\eea

The FHM Hamiltonian preserves numbers of electrons with spins $\uparrow$ and $\downarrow$. Therefore,  $\rho(\beta)$ is invariant under a symmetry transformation $T$, 
\begin{equation}
T \rho(\beta) T^{\dag} = \rho(\beta), 
\end{equation}
where $T= \otimes_j T^{ph}_j$ is a product over all lattice sites,  $T^{ph}_j = T_j^{\uparrow} \otimes T_j^{\downarrow}$ is a unitary matrix representation of $U(1)\times U(1)$ group with  $T_j^{m}  = e^{-i \phi n_j^{m}}$, where $n_j^m$ is a particle number operator with spin $m$ on site $j$, and $\phi \in \mathbb{R}$.

\begin{figure*}[t!]
\vspace{-0cm}
\includegraphics[width=0.55\textwidth,clip=true]{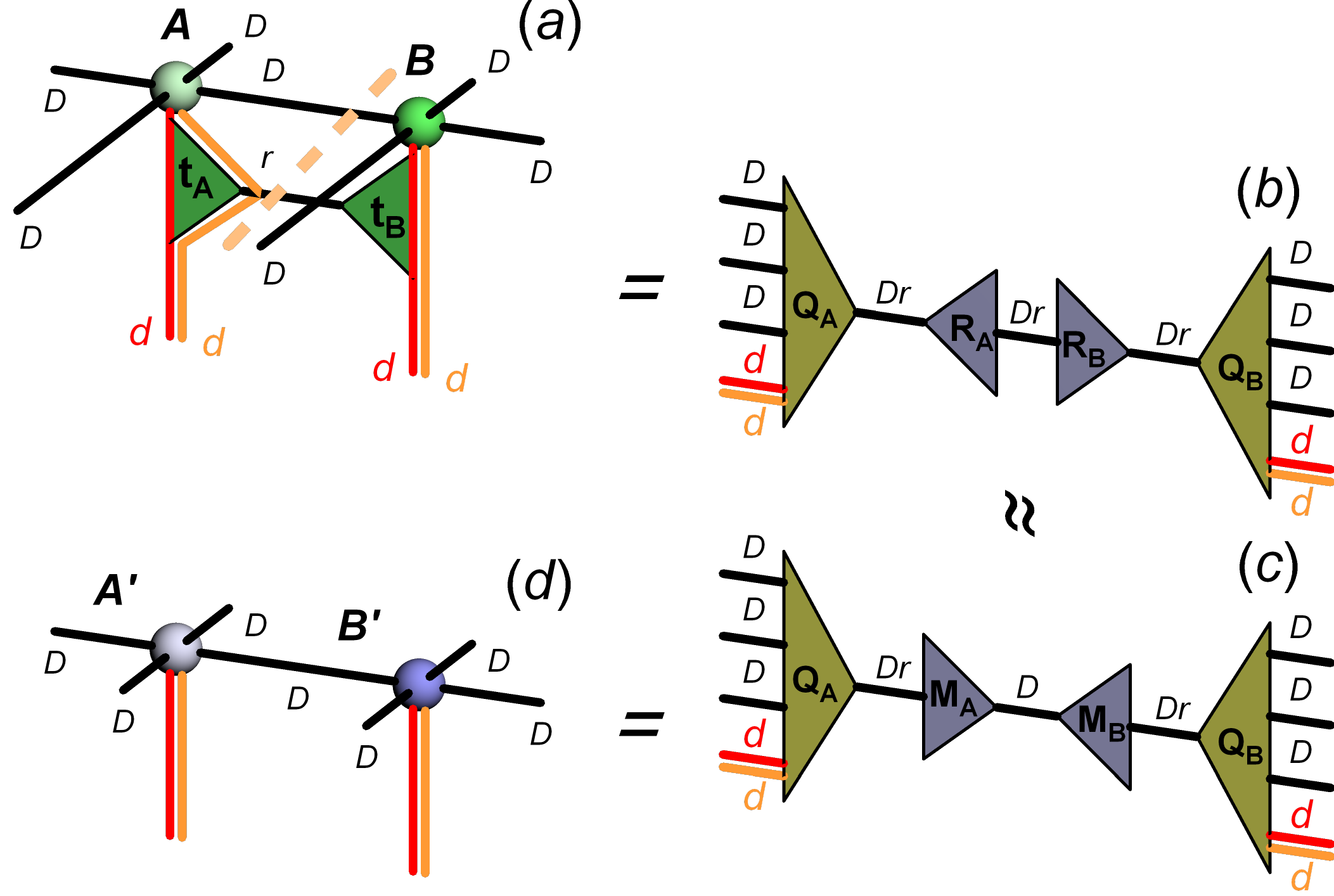}
\vspace{-0cm}
\caption{
{\bf Application of the Trotter gate. }
In (a), 
a 2-site gate is applied to nearest neighbor tensors $A$ and $B$ as in Fig.~\ref{fig:NTU_overview}(b). 
The Trotter gate is a contraction of two tensors, $t_A$ and $t_B$, through an index with dimension $r$.
In (b),
the tensor contraction $A\cdot t_A$ is QR-decomposed into $Q_A \cdot R_A$. Similarly, $B\cdot t_B=Q_B \cdot R_B$. 
Isometries $Q_{A}$ and $Q_B$ remain fixed during the optimization procedure.
In (c), the exact product $R_A R_B^T$ is approximated by $M_A M_B^T$ with the original bond dimension $D$.
This can be done either through a simple SVD or the environment-assisted truncation (EAT), see App.~\ref{app:EAT}.
In (d), new iPEPS tensors could be obtained as $A'=Q_A\cdot M_A$ and $B'=Q_B\cdot M_B$ completing the Trotter gate. First, however, matrices $M_{A}$ and $M_{B}$ are optimized in the neighborhood tensor environment, as shown in Figs.~\ref{fig:NTUenv} and~\ref{fig:NTUloop}, before being contracted back with $Q_{A}$ and $Q_B$ to form new iPEPS tensors $A'$ and $B'$. 
}
\label{fig:2site2}
\end{figure*}

To enforce this abelian symmetry, an iPEPS representing $\rho(\beta)$ is constructed from $U(1)\times U(1)$ invariant tensors~\cite{singh2010,bauer2011}, 
\begin{equation}
A_{satlbr} = \sum_{s'a't'l'b'r'}  T^{ph}_{s s'} T^{ph\dag}_{a a'}  T^{(t)}_{t t'}  T^{(l)\dag}_{l l'} T^{(b)\dag}_{b b'} T^{(r)}_{r r'} A_{s'a't'l'b'r'}, 
\label{eq:AT}
\end{equation}
and analogously for $B$. Here,  $T^{(t)}$, $T^{(l)}$, $T^{(b)}$ and $T^{(r)}$ are $U(1)\times U(1)$  unitary matrix representations acting at virtual indices of the iPEPS tensor $A$, and $s$ and $a$, respectively, label physical and ancillary degrees of freedom of a lattice site. It can now be decomposed into symmetric sectors labeled by charges ${\bf t}_{s}, {\bf t}_{a}, {\bf t}_{t}, {\bf t}_{l}, {\bf t}_{b}, {\bf t}_{r}$ corresponding to each index of $A$~\cite{singh2010}, 
\begin{equation}
A = \bigoplus{}_{{\bf t}_{s}, {\bf t}_{a}, {\bf t}_{t}, {\bf t}_{l}, {\bf t}_{b}, {\bf t}_{r}  } A^{{\bf t}_{s}, {\bf t}_{a}, {\bf t}_{t}, {\bf t}_{l}, {\bf t}_{b}, {\bf t}_{r}}. \label{eq:Ablock}
\end{equation}Dimensions of virtual indices of the sectors are called sectorial bond dimensions $D_{\bf t}$. 

In the case of $U(1) \times U(1)$, the charges are formed by pairs of integers, ${\bf{t}} = (t^\uparrow, t^\downarrow)$. To ensure $U(1)\times U(1)$ invariance of $A$ and $B$, they obey
\be
t_{s}^m - t_{a}^m + t_{t}^m - t_{l}^m - t_{b}^m + t_{r}^m=0, \quad m = \uparrow,\downarrow, 
\label{eq:charge_condition}
\ee
where the signs (or signatures) correspond to hermitian conjugations in Eq.~\eqref{eq:AT}.  For non-interacting spinless fermions discussed in App.~\ref{app:benchmarks}, we have preservation of the total number of fermions manifesting itself as $U(1)$ symmetry of iPEPS tensors.  In such a case  $T^{ph}_j = e^{-i \phi n_j}$, where $n_j$ is a fermion density operator at site $j$ and the charges are integers ${\bf t}_{s}, {\bf t}_{a}, {\bf t}_{t}, {\bf t}_{l}, {\bf t}_{b}, {\bf t}_{r}$ summing up to zero as in Eq.~\eqref{eq:charge_condition}. 

The symmetries are implemented with YAST symmetric tensor library~\cite{YAST} that we employ in this work. They are instrumental not only to obtain sparser tensors, allowing to reach significantly larger bond dimensions $D$, but also to enforce fermionic statistics. For the latter, we follow a general scheme of Refs.~\cite{Corboz09_fmera,Corboz_fiPEPS_10}. 

Enforcing fermionic statistics amounts to projecting the tensor network on a plain where the line crossings indicate the application of a {\it SWAP gate}. This requires symmetric tensors with fermionic parity defined for each tensor leg. In our case, the fermionic parity equals parity of $t^{\uparrow}+t^{\downarrow}$. A SWAP gate applied to two tensor legs multiplies by $-1$ all tensor blocks (defined in Eq.~\eqref{eq:Ablock} for a particular tensor with 6 legs) that have odd fermionic parity on both those legs. In particular, such leg crossings appear in Figs.~\ref{fig:NTU_overview}, \ref{fig:2site2}, and~\ref{fig:NTUenv}. These figures are the building blocks of the NTU algorithm described in the next section. 

All expectation values in this work are calculated using the standard corner transfer matrix renormalization group (CTMRG) ~\cite{nishino1996,Orus_CTM_09,Corboz_CTM_14} and have been converged against environmental bond dimension $\chi$ which is its refinement parameter. 
CTMRG is also used to estimate the largest correlation length $\xi$ in the system using the largest eigenvalues of  CTMRG  row-to-row and column-to-column transfer matrices~\cite{nishino96}. We ensured that   they have been converged against $\chi$ as well.

\section{NTU evolution algorithm}
\label{app:NTU}

\begin{figure*}[t!]
\vspace{-0cm}
\includegraphics[width=0.7\textwidth,clip=true]{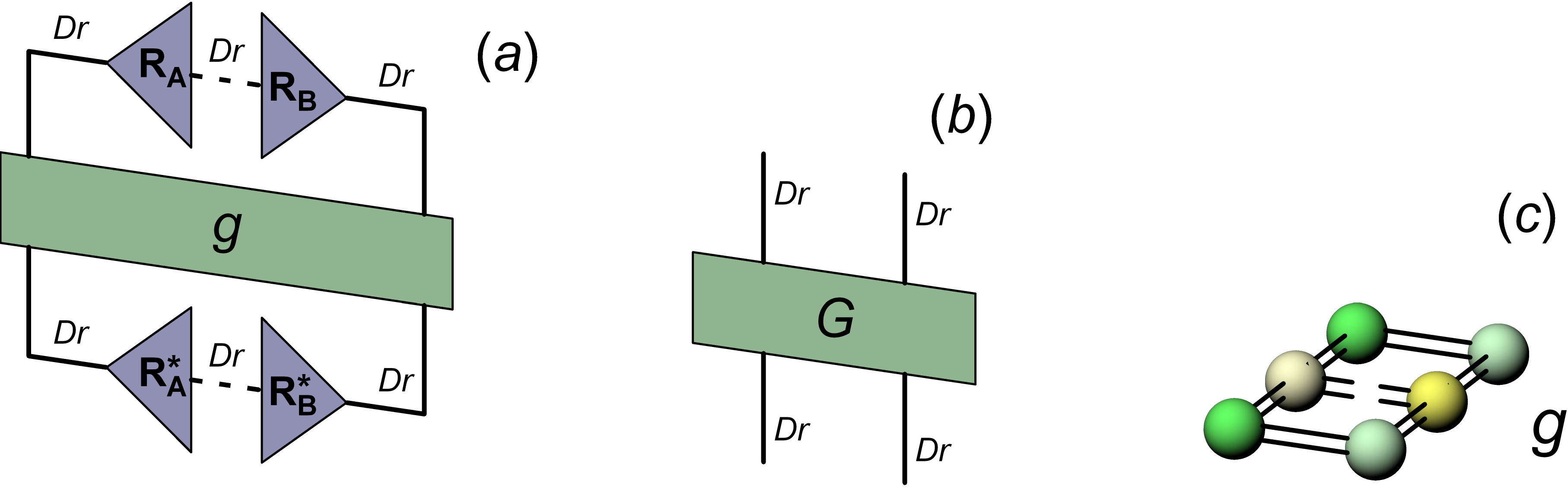}
\includegraphics[width=0.7\textwidth,clip=true]{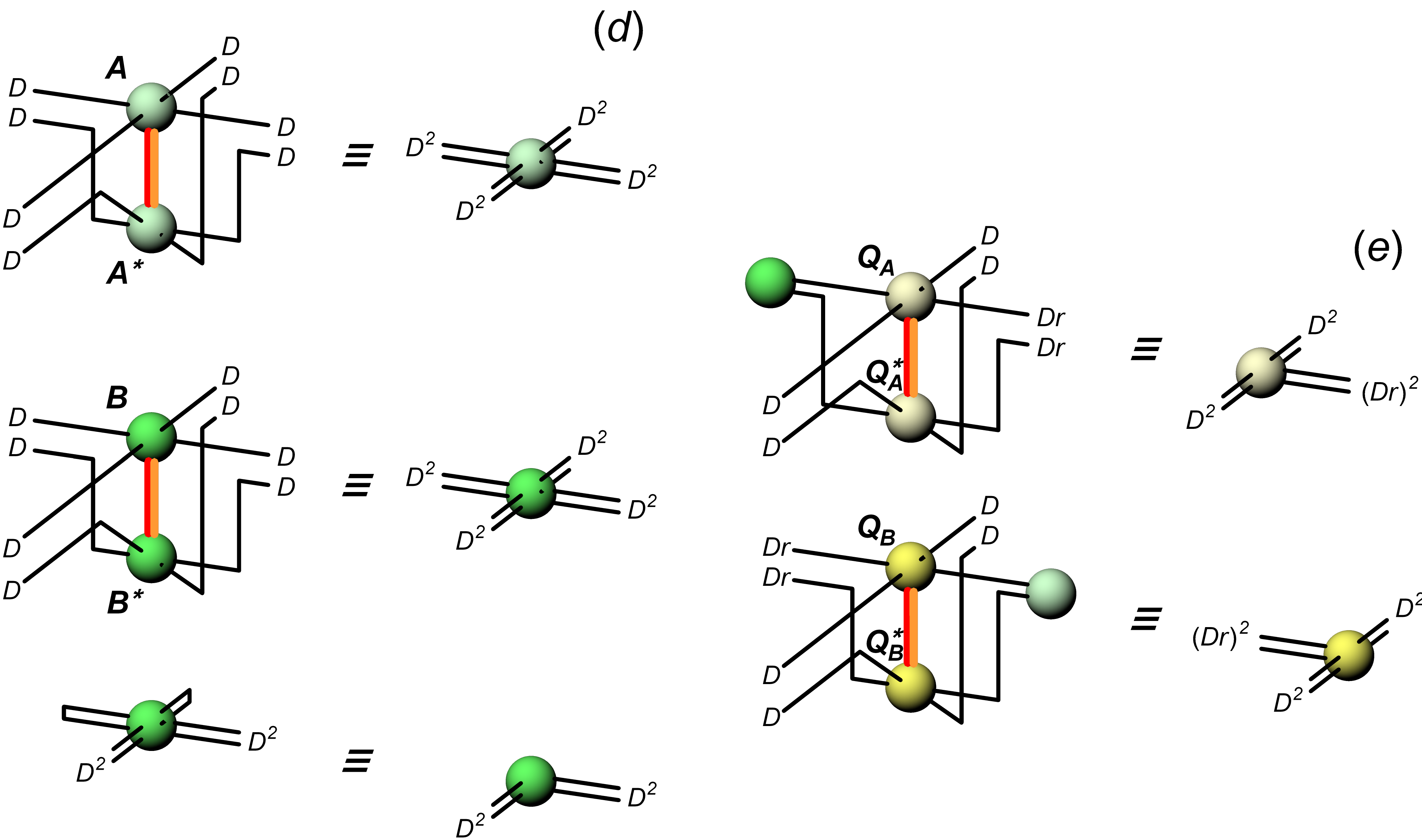}
\vspace{-0cm}
\caption{
{\bf NTU metric tensor. }
Norm squared of the matrix product, $\vert\vert R_AR_B^T \vert\vert^2$, calculated in the environment of the NTU cluster from Fig.~\ref{fig:NTU_overview}(b) is shown in (a).
Here $g$ is a metric tensor assembled in  (c). 
The upper (lower) pair of free indices in (c) corresponds to the upper (lower) pair of indices of $g$ in (a). 
The diagram in (c) is a contraction of 6 tensors. Three of them are shown in (d) and (e), while the remaining three are constructed similarly to the one in (d). In (d) we also show double iPEPS tensors that appear in the norm squared of the iPEPS, $\left<\psi|\psi\right>$. For a fermionic iPEPS, any crossing of two lines implies a SWAP gate. Note that the enforced tensor block structure, such as in Eq.~\eqref{eq:charge_condition}, allows one to pull lines over the tensors changing the placement of SWAP gates without changing the overall result~\cite{Corboz09_fmera,Corboz_fiPEPS_10}. For that reason, all
the necessary SWAP gates can be applied with sub-leading computational cost, such as in panel (d).
Finally, when both dashed bonds in (a) are cut, we obtain bond metric tensor $G$ in (b).
It is the starting point for EAT in Fig.~\ref{fig:EAT} and App.~\ref{app:EAT}. 
}
\label{fig:NTUenv}
\end{figure*}

\begin{figure*}[t!]
\vspace{-0cm}
\includegraphics[width=0.5\textwidth,clip=true]{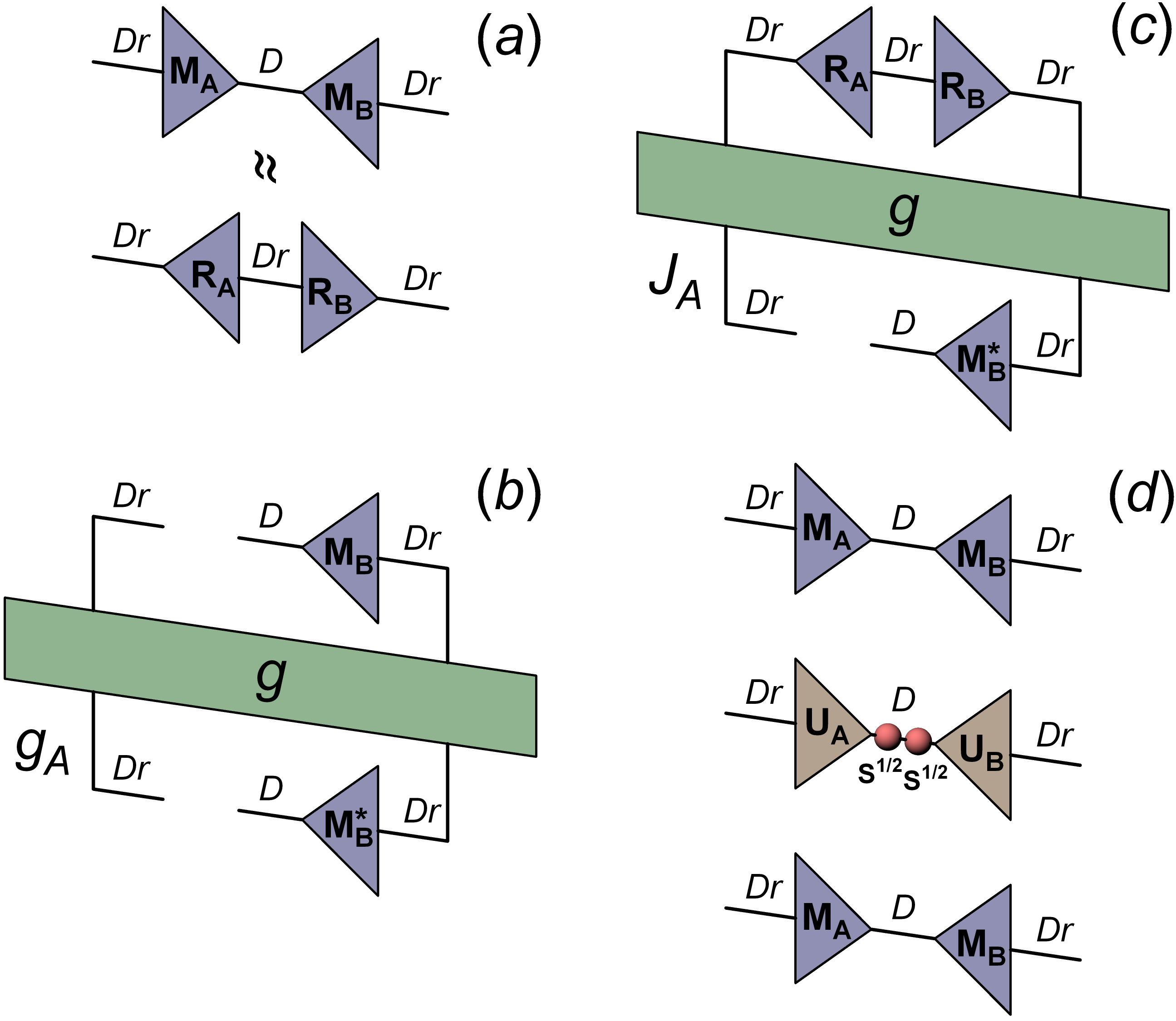}
\vspace{-0cm}
\caption{
{\bf Optimization of reduced matrices. }
In (a), matrices $M_A$ and $M_B$ are optimized for their product, $M_A M_B^T$, to be the best approximation to the exact product, $R_AR_B^T$. The error is measured with respect to the metric tensor $g$ shown in Fig.~\ref{fig:NTUenv}(a). 
In (b), we depict the reduced metric tensor $g_A$ for matrix $M_A$, and
in (c) the reduced term $J_A$. 
In (d), the product of the matrices is subject to SVD, $M_A M_B=U_ASU_B^T$.
Finally, when the matrices are converged, new balanced matrices $M_A=U_AS^{1/2}$ and $M_B^T=S^{1/2}U_B^T$ are formed by symmetrically absorbing singular values $S$. 
However, during the iterative optimization, they are not kept balanced. Before optimization with respect to $M_A$ the matrices are ``tilted'' as $M_A=U_AS$ and $M_B^T=U_B^T$ and vice versa~\cite{Evenbly2018}. This way, the optimized $M_A$ represents larger chunk of the product $M_AM_B^T$, whose optimization is the ultimate goal of the iterative procedure.  
}
\label{fig:NTUloop}
\end{figure*}

The time-evolution method is explained in some detail by the diagrams in Figs.~\ref{fig:NTU_overview},~\ref{fig:2site2},~\ref{fig:NTUenv}, and~\ref{fig:NTUloop}. The following text serves as a guide through the figures. 

The evolution operator $U(\beta)$ is applied to the tensors sequentially as a product of small time-steps $U(d\beta)$, each of them approximated by a series of local gates via the second-order Suzuki-Trotter decomposition.  Fig.~\ref{fig:2site2}(a) shows in detail the gate applied to a horizontal pair of nearest-neighbor iPEPS tensors, $A$ and $B$, the same as in Fig.~\ref{fig:NTU_overview}(b). The rank-$r$ gate enlarges the bond dimension from $D$ to $r \cdot D$, that will be truncated back to $D$. For better numerical efficiency, we  use QR decomposition to compute reduced matrices $R_A$ and $R_B$ in place of full tensors~\cite{Evenbly2018}, see Fig.~\ref{fig:2site2}(b). Their product, $R_AR_B^T$, is to be approximated by a product of new matrices contracted through a bond of dimension $D$: $M_AM_B^T$, see Fig.~\ref{fig:2site2}(c). Those matrices get combined with isometries $Q_A$ and $Q_B$ into the new iPEPS tensors $A'$ and $B'$ in Fig.~\ref{fig:2site2}(d). However, before this final contraction, the $M$ matrices are subject to NTU optimization.

The NTU optimization of the reduced matrices $M_A$ and $M_B$ minimizes the Frobenius norm of the difference between the two sides of the equation in Fig.~\ref{fig:NTU_overview}(b). 
As the diagrams in Figs.~\ref{fig:2site2}(c) and (d) are equal, the RHS of the equation in Fig.~\ref{fig:NTU_overview}(b) is linear in product $M_AM_B^T$, and the norm squared of the difference between the LHS and the RHS can be written as
\be
F{\left[M_AM_B^T\right]}=
\left(M_AM_B^T-R_AR_B^T\right)^\dag ~ g ~ \left(M_AM_B^T-R_AR_B^T\right),
\label{cost}
\ee 
where $g$ is a metric tensor defined by this construction. The tensor can be directly computed as in Fig.~\ref{fig:NTUenv}. Thanks to its numerical exactness, $g$ is a manifestly non-negative and Hermitian matrix. Matrices $M_A$ and $M_B$ are optimized to minimize the cost function~\eqref{cost} or, equivalently, to make their product, $M_AM_B^T$, the best approximation to the exact product, $R_AR_B^T$. Their optimization proceeds iteratively:
\be 
\ldots \to M_A \to M_B \to M_A \to M_B \to \ldots
\label{eq:iteration}
\ee 
until convergence of the cost function.

When optimizing $M_A$ for fixed $M_B$, the cost function~\eqref{cost} becomes quadratic in $M_A$:
\be 
F_A{\left[M_A\right]} = M_A^\dag g_A M_A - M_A^\dag J_A - J_A^\dag M_A + F_0.
\ee 
Here $g_A$ and $J_A$ depend on $M_B$, see Figs.~\ref{fig:NTUloop}(b) and (c), and $M_{A/B}$-independent $F_0$ is shown in Fig.~\ref{fig:NTUenv}(a). The matrix is updated as
\be 
M_A={\rm pinv}\left(g_A\right)J_A,
\ee 
where the tolerance of the pseudo-inverse is dynamically adjusted to minimize $F_A{\left[{\rm pinv}\left(g_A\right)J_A\right]}$. Thanks to the exactness of $g$, the reduced $g_A$ is also a manifestly non-negative and Hermitian matrix. As there is no need to correct exact $g_A$, the only role of the dynamical tolerance is to keep the influence of numerical inversion errors under control. In practice, the optimal tolerance remains in the range $10^{-12}$---$10^{-8}$ relative to the maximal eigenvalue of $g_A$. The numerical exactness of $g_A$ and its practical consequences provide key motivation behind the NTU scheme.

The optimization of $M_A$ is followed by a similar optimization of $M_B$. The two optimizations are repeated until convergence of a relative NTU error: 
\be 
\delta = \sqrt{ F{\left[M_AM_B^T\right]}/F_0} . 
\label{NTUerror}
\ee 
This error measures the accuracy of the NTU tensor truncation. The square root makes $\delta$ an estimate for a relative error of the purification inflicted by the truncation after the Trotter gate and thus also of errors of its expectation values. Therefore, for a small enough imaginary time-step, it should become proportional to $d\beta$  making $\delta / d\beta$ a step-size-independent measure of the error caused by the truncation. 

The truncation errors accumulate with evolution time. As long as the errors remain small, the worst case scenario is that they are additive. The additiveness should hold at least over short time intervals over which the purification does not change much, and small errors made by subsequent truncations point in approximately the same direction in the Hilbert space. This heuristic reasoning motivates an integrated NTU error:
\be 
\Delta \equiv \sum_i \delta_i,
\label{sigmadelta}
\ee 
where the sum is over all Trotter gates between $0$ and $\beta$, as a relevant estimate of the purification error at $\beta$. We employ this estimate in the main text.

\begin{figure*}[t!]
\vspace{-0cm}
\includegraphics[width=0.40\textwidth,clip=true]{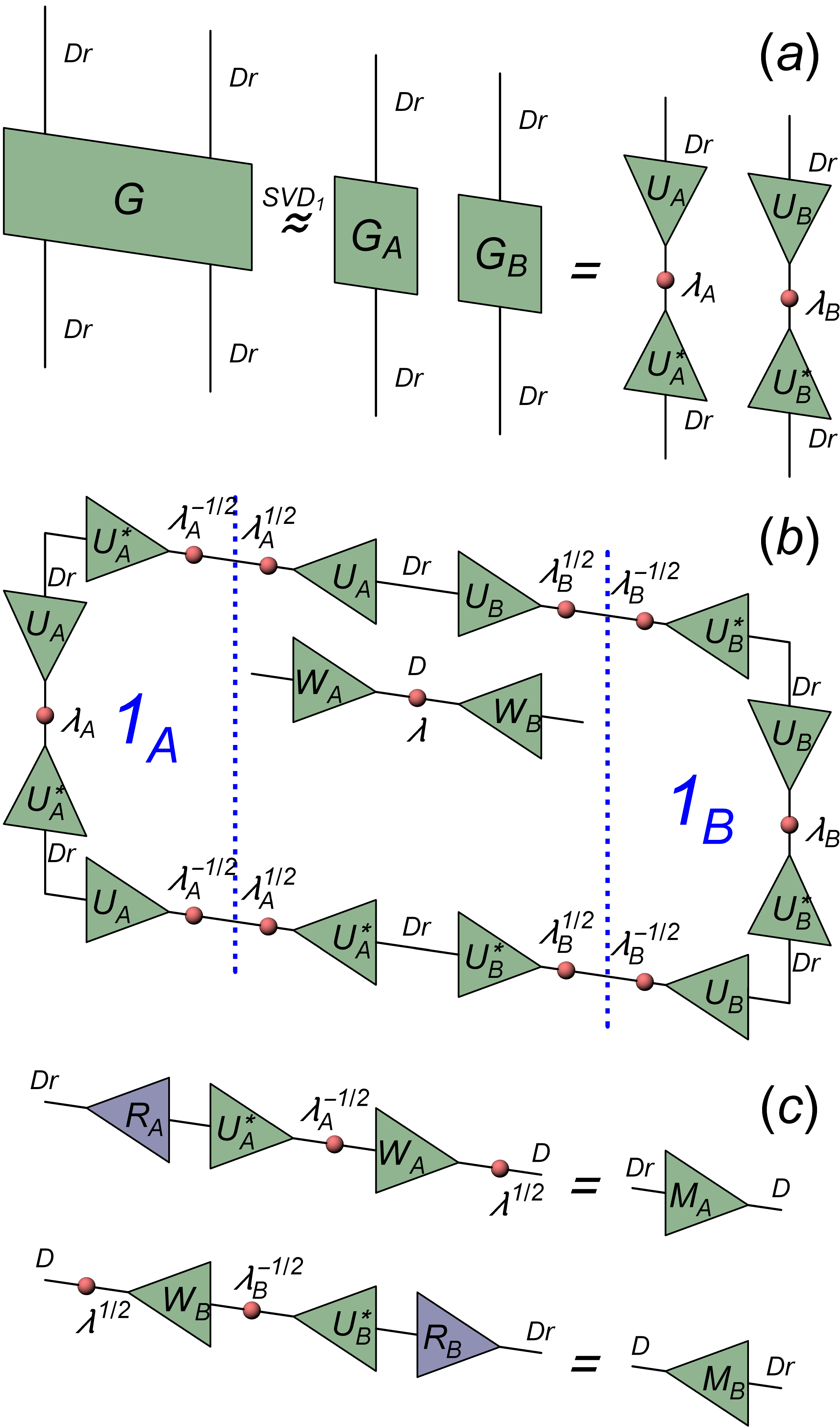}
\vspace{-0cm}
\caption{
{\bf Environment-assisted truncation (EAT). }
In (a), 
the NTU metric tensor defined in Fig.~\ref{fig:NTUenv}(b) is approximated by a product of two metric tensors, $G\approx G_A\otimes G_B$.  The approximation is made by SVD truncated to its leading singular value. Each of them is diagonalized as, e.g., $G_A=U_A\lambda_AU_A^\dag$, where $\lambda_A\geq0$.
In (b), the top and bottom pairs of indices of the product $G_A\otimes G_B$ are inserted with identities. This gauge transformation makes the product metric tensor an identity, $1_A\otimes 1_B$, with matrix $\lambda_A^{1/2}U_A^TU_B\lambda_B^{1/2}\equiv m$ inserted in its top and bottom indices. In this identity metric, SVD provides the optimal way to truncate the bond dimension: $m=W_A\lambda W_B^T$.  
In (c), the SVD matrices are used to initialize reduced matrices $M_A$ and $M_B$ in Fig.~\ref{fig:2site2}(c). 
}
\label{fig:EAT}
\end{figure*}

\section{Initialization of updated tensors}
\label{app:EAT}

In this section, we elaborate on the initialization of matrices $M_A$ and $M_B$,  see Fig.~\ref{fig:2site2}(b) and (c). As already indicated in the figure caption, a traditional strategy, practiced in full update, see~\cite{bruognolo2021beginner} for an introductory review, is to make an SVD decomposition of the product $R_AR_B^T$ before truncating it to $D$ dominant singular values. 

The state-of-the-art SVD initialization, however, knows nothing about the tensor environment of the product. With poor initialization, the iteration procedure of Eq.~\eqref{eq:iteration} may end up getting trapped in a local minimum. More importantly, for the symmetric iPEPS, the total bond dimension $D$ is a sum of sectorial bond dimensions $D_{\bf{t}}$. The iterative NTU optimization~\eqref{eq:iteration} is working within a fixed distribution of $D_{\bf{t}}$, not being able to update it even though it takes into account the NTU environment exactly. This might lead to misrepresentation of the evolved state and makes proper initialization of truncated matrices $M_A$ and $M_B$, with a particular distribution of $D_{\bf{t}}$, a crucial part of a successful algorithm. Below, we discuss two methods that we employ in this work.

\subsection{Environment-assisted truncation}

In order to take into account the environment in an approximate way, we propose the environment-assisted truncation (EAT). While here the environment taken into account is the NTU cluster in Fig.~\ref{fig:NTU_overview}(b), the scheme can be directly employed in larger or infinite environments.

The norm squared of the product $R_AR_B^T$ is shown in Fig.~\ref{fig:NTUenv}(a), where the metric tensor $g$ encapsulates relevant information about the NTU environment. 
Cutting the dashed bonds in Fig.~\ref{fig:NTUenv}(a) creates metric tensor $G$ in Fig.~\ref{fig:NTUenv}(b). 
By construction, within the NTU scheme, it is manifestly Hermitian and non-negative. 
If an object were inserted in the dashed bonds then $G$ would measure its norm.
If the object were a projector truncating the bond dimension then $G$ would measure the error of the truncation.

EAT approximates $G$ in Fig.~\ref{fig:NTUenv}(b) with a product of metric tensors, $G_A \otimes G_B$, see Fig.~\ref{fig:EAT}. The approximation is done by an SVD of $G$ between its left and right indices, truncated to the dominant singular value. After eventual adjustment of phases of the leading left and right singular vectors, both $G_A$ and $G_B$ are manifestly Hermitian and non-negative. This property is inherited from the NTU metric $g$. The advantage of the product is that -- while it does not ignore the tensor environment -- reduced matrices $M_A$ and $M_B$ can still be initialized using a simple SVD, see Fig.~\ref{fig:EAT}(b). Similarly as after the traditional SVD initialization, the initial matrices $M_A$ and $M_B$ are further optimized to minimize the NTU error in the exact NTU environment.  

Before we proceed, to put it into broader context, it is worth considering the application of EAT in a one-dimensional setup of matrix product states. In that case, the rank-1 approximation of the metric tensor performed in Fig.~\ref{fig:EAT}(a) would be exact, with $G_A$ and $G_B$ being the exact left and right environments of a given bond. The following steps in Fig.~\ref{fig:EAT} amount in that case to the optimal truncation of that bond~\cite{Orus2008}, without the need for further iterative updates.
This is not the case for a 2D setup, where EAT in~\ref{fig:EAT}(a) takes the first product approximation of the environment, but still going beyond the standard SVD initialization that does not input any information about the environment.

\begin{table*}
\centering
\begin{tabular}{|| c | c ||} 
\hline
$D$ & \{${\bf{t}}:D_{\bf{t}}$\} \\  [0.5ex] 
\hline\hline
 14 & {(0,0):2, (-1,0):2, (1,0):2, (0,-1):2, (0,1):2, (-1,-1):1, (-1,1):1, (1,-1):1, (1,1):1} \\
 \hline
 15 & {(0,0):3, (-1,0):2, (1,0):2, (0,-1):2, (0,1):2, (-1,-1):1, (-1,1):1, (1,-1):1, (1,1):1}\\
 \hline
 16 & {(0,0):4, (-1,0):2, (1,0):2, (0,-1):2, (0,1):2, (-1,-1):1, (-1,1):1, (1,-1):1, (1,1):1} \\
 \hline
 20 & {(0,0):4, (-1,0):2, (1,0):2, (0,-1):2, (0,1):2, (-1,-1):2, (-1,1):2, (1,-1):2, (1,1):2} \\
 \hline
 24 & {(0,0):4, (-1,0):3, (1,0):3, (0,-1):3, (0,1):3, (-1,-1):2, (-1,1):2, (1,-1):2, (1,1):2} \\
 \hline
 25 & {(0,0):5, (-1,0):3, (1,0):3, (0,-1):3, (0,1):3, (-1,-1):2, (-1,1):2, (1,-1):2, (1,1):2}  \\ 
 \hline
 26 & {(0,0):6, (-1,0):3, (1,0):3, (0,-1):3, (0,1):3, (-1,-1):2, (-1,1):2, (1,-1):2, (1,1):2}  \\
 \hline
 29 & {(0,0):5, (-1,0):4, (1,0):4, (0,-1):4, (0,1):4, (-1,-1):2, (-1,1):2, (1,-1):2, (1,1):2}  \\
 \hline
\end{tabular}
\hspace{0.6cm}
\begin{tabular}{|| c | c ||} 
\hline
$D$ & \{${\bf{t}}:D_{\bf{t}}$\}\\  [0.5ex] 
\hline\hline
 7 & {-1:2, 0:3, 1:2} \\ 
 \hline
 9 & {-2:1, -1:2, 0:3, 1:2, 2:1}\\
 \hline
 10 & {-2:1, -1:2, 0:4, 1:2, 2:1} \\
 \hline
 11 & {-2:1, -1:2, 0:5, 1:2, 2:1} \\
 \hline
 14 &  {-2:1, -1:4, 0:4, 1:4, 2:1}  \\ 
 \hline
 16 &  {-2:1, -1:3, 0:8, 1:3, 2:1}  \\ 
 \hline
 19 &  {-2:2, -1:4, 0:7, 1:4, 2:2}  \\ 
 \hline
 25 &  {-2:2, -1:6, 0:9, 1:6, 2:2} \\ 
\hline
\end{tabular}
\caption{
{\bf Fixed distributions of sectorial bond dimensions.} In the left table, we list charges and the corresponding sectorial bond dimensions of virtual legs of  $U(1)\times U(1)$  symmetric iPEPS for various total bond dimensions $D$ used in the simulations of the Hubbard model within the FIX initialization scheme. In the right table, we show the corresponding data for $U(1)$ symmetric iPEPS employed in benchmarks for a non-interacting fermion model in App.~\ref{app:benchmarks}.}
\label{table:dt}
\end{table*}

One may now consider options to further improve the initialization procedure. One such option is to perform a gradual truncation within a $m$-step EAT+NTU (EATm). In this procedure the initial truncation is done not in one but in $m$ steps as $rD\to D_{m-1} \to \dots \to D_1 \to D$, where $r$ is the SVD rank of the Trotter gate applied. For example, a 2-step EAT (EAT2) would involve truncating $rD$ to, say, $rD/2$ with EAT, followed by optimization with the NTU metric, and then subsequent truncation from  $rD/2$ to $D$, ending with a final NTU optimization. 
During the first truncation metric $G$ is subject to the product approximation, $G\approx G_A\otimes G_B$, making the partial truncation of matrices $M_A$ and $M_B$ suboptimal. The following NTU optimization improves the matrices with respect to exact metric $g$. This improvement is reflected in a new metric $G$ --- constructed as in Figs.~\ref{fig:NTUenv}(a,b) but with the partly truncated $M_A$ and $M_B$ in place of the untruncated $R_A$ and $R_B$ --- defining the error measure for the second truncation. Consequently, the 2-step EAT should be less affected by the product approximation and, in particular, provide better choice of sectorial bond dimensions $D_{\bf{t}}$ than the 1-step EAT.

\begin{figure}[b!]
\vspace{-0cm}
\includegraphics[width=\columnwidth,clip=true]{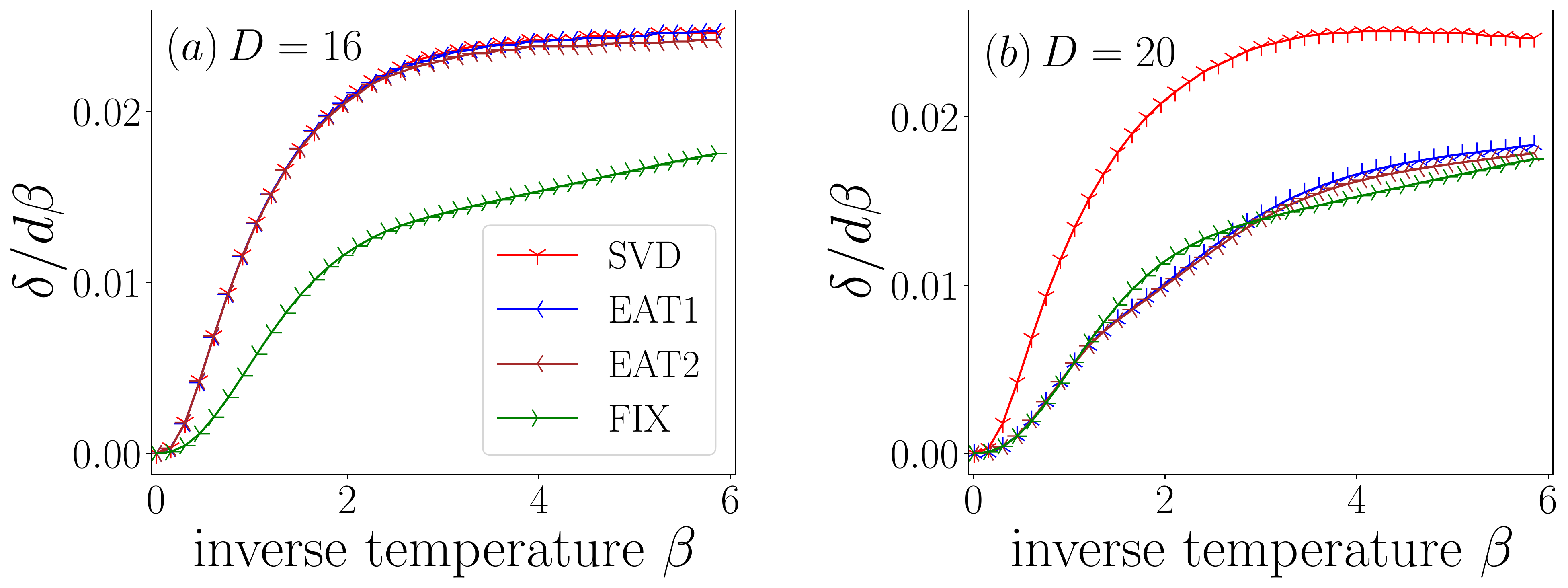}
\vspace{-0cm}
\caption{
{\bf Comparison of performance  of NTU initialization schemes. }
The NTU error $\delta$ normalized by the Trotter step $d\beta$ for total bond dimensions $D=16$ (a) and $D=20$ (b) plotted versus $\beta$ for   imaginary time evolution of the Hubbard model at half-filling. Different curves correspond to different NTU optimization initialization strategies, SVD, EAT1, EAT2, and FIX. We plot here $\delta$ for final NTU optimized tensors demonstrating their strong dependence on the initialization. We find that the proposed EAT1, EAT2, and FIX schemes improve over the standard SVD approach. }
\label{fig:whyfix}
\end{figure}

\begin{figure*}[t!]
\vspace{-0cm}
\includegraphics[width=\textwidth,clip=true]{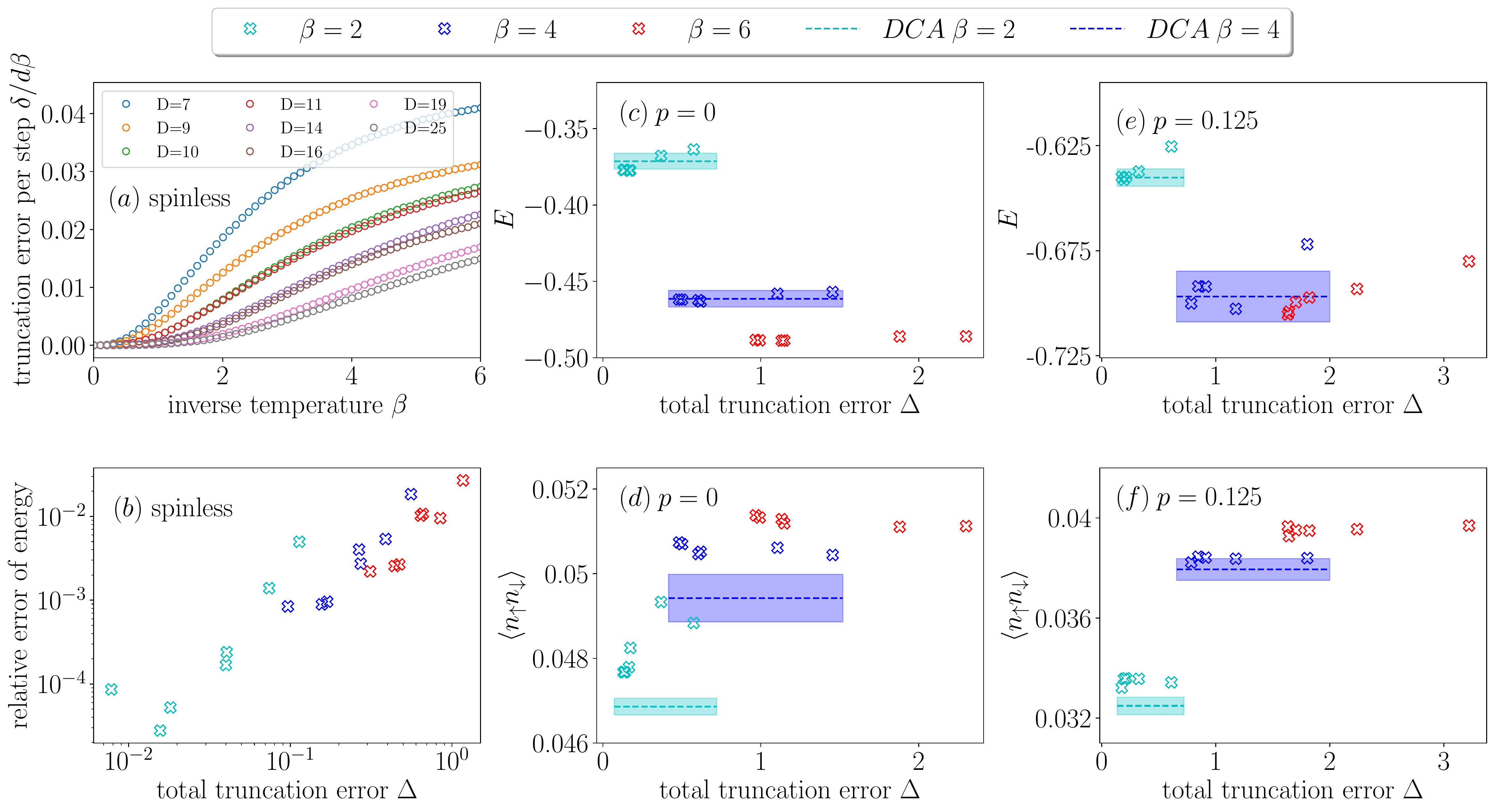}
\vspace{-0cm}
\caption{
{\bf Benchmarks. } 
In (a), we plot the NTU error~\eqref{NTUerror} versus inverse temperature $\beta$ for the imaginary time evolution of an analytically solvable spinless fermions model (\ref{H0}).  Here, the error is averaged over all Trotter gates in each time step, and $D$ is ranging from $7$ to $25$. 
In (b), we plot the absolute relative error for energy in the function of the total truncation error $\Delta$ [integrated NTU error in Eq.~\eqref{sigmadelta}] for  the spinless fermions and $\beta=2,4,6$.
Next we consider  FHM with $U=8$ and $\beta = 2,4,6$. 
 In (c) and (d), we show energy $E$ and double occupancy $\left<n_{\uparrow} n_{\downarrow}\right>$ versus $\Delta$  for $p=0$.  In (e) and (f), we show the same for $p=0.125$. We perform simulations with the FIX initialization scheme as described in App.~\ref{app:EAT}, for the total bond dimensions in $14 \le D \le 29$. For the comparison, in (c)-(f), we  show DCA results~\cite{leblanc15}  marked with dashed lines surrounded by contours indicating the results' uncertainty. In our simulations we preserve doping $p=0$ within accuracy of $10^{-6}$  and $p=0.125$ within accuracy of $5\times10^{-4}$.}
\label{fig:benchmarks}
\end{figure*}

\subsection{Fixed distribution}

Another truncation strategy is to constrain $D_{\bf{t}}$ by hand in a way that reflects the spatial symmetries of the problem. First, we choose the same set of charges $\bf{t}$ and their respective bond dimensions $D_{\bf{t}}$ for each virtual leg of the iPEPS tensors. Second, we assign to charges with opposite signs, like $(1,-1)$ and $(-1,1)$, the same bond dimensions, guided by an intuition that a current of fermions along a bond should be zero. The same set of $D_{\bf{t}}$ is used throughout the whole imaginary time evolution. We try different distributions obeying those constraints and accept the one that yields the minimal NTU error. We call this strategy FIX. 
 We collect virtual leg charges ${\bf{t}}$ and their respective bond dimensions $D_{\bf{t}}$ found with FIX in Tab.~\ref{table:dt}. Results in the main text and following App.~\ref{app:benchmarks} have been obtained with the parameters listed in the tables. 

Initialization of matrices $M_A$ and $M_B$ in each charge sector with predefined $D_{\bf{t}}$ was performed with EAT. It gives a better initial NTU error, $\delta$, than the SVD truncation and, therefore, should help to prevent the following NTU optimization of $\delta$ from getting trapped in a local minimum. For the half-filled Hubbard model we find that using FIX with EAT initialization typically results in the best NTU error while comparing to SVD and EAT schemes, see examples in Fig.~\ref{fig:whyfix}. We note  that  for some $D$, EAT or both SVD and EAT initialization give similar $\delta$ as FIX. This behavior is not unexpected as some optimization instances can be less affected by local minima than others. Consequently, we use FIX with EAT initialization (the combo being labelled as FIX for simplicity---which has negligible overhead over FIX with SVD initialization) for all the simulations in the main text and the following benchmarks in App.~\ref{app:benchmarks}. 

We compare the final NTU errors for evolutions with the EAT initialization schemes and SVD scheme for the Hubbard model at half-filling using the NTU error $\delta$. 
We see that in some cases both 1-step and 2-step EAT clearly outperforms the SVD initialization while in the others they gives results of similar quality, see examples in Fig.~\ref{fig:whyfix}. %
In our simulations we find that a 2-step EAT (EAT2) procedure, in general, leads to slightly lower NTU error than the 1-step version (EAT1), and we use the former for our benchmarks in App.~\ref{app:benchmarks}. For simplicity, we henceforth label a 2-step EAT initialization procedure as EAT.

\begin{figure}[b!]
\vspace{-0cm}
\includegraphics[width=\columnwidth,clip=true]{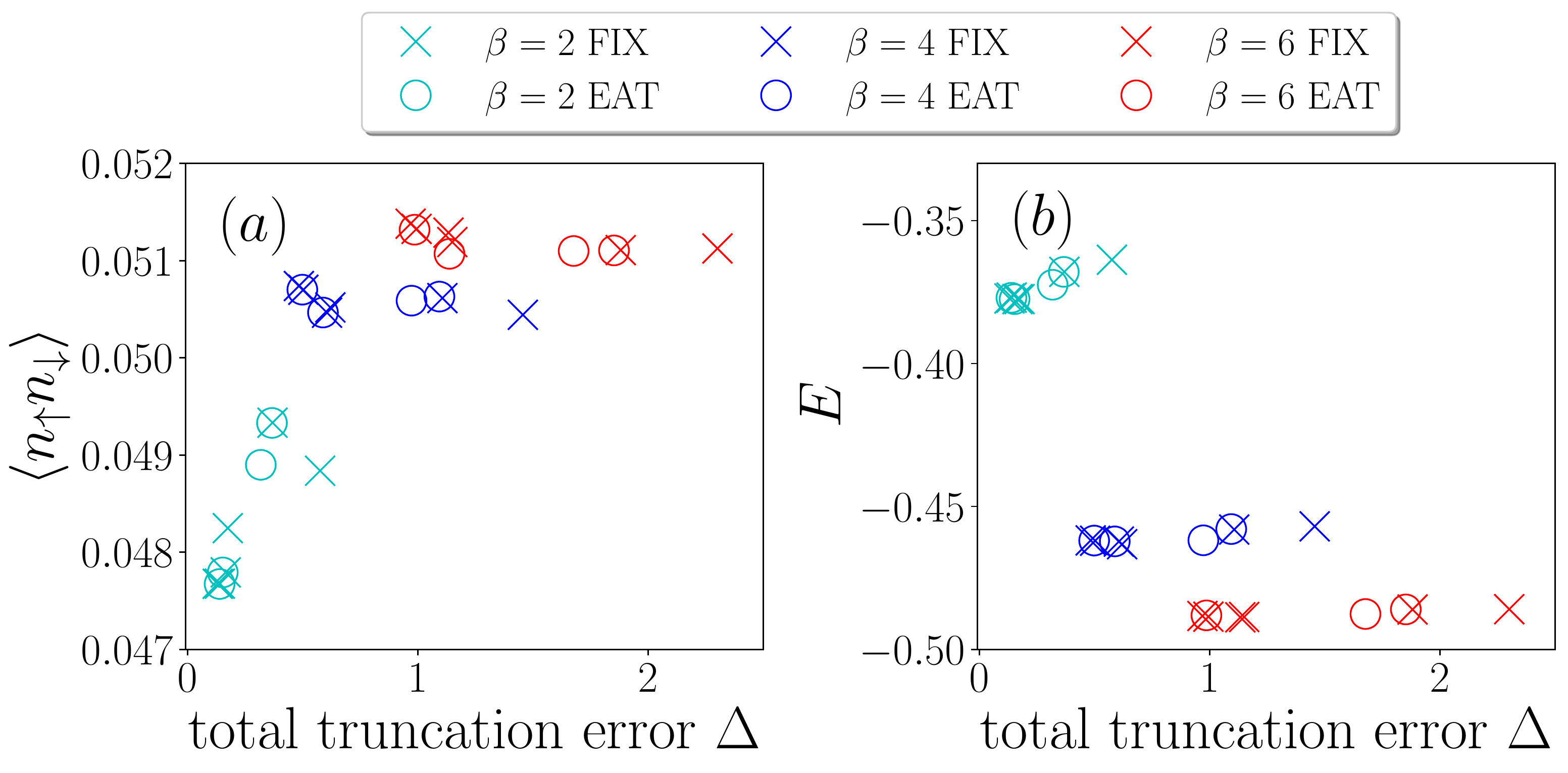}
\vspace{-0cm}
\caption{
{\bf Consistency of the results for FIX and EAT initialization schemes.} 
A comparison between the expectation values obtained from the two initialization strategies introduced in App.~\ref{app:EAT}. In (a), we plot the energy $E$ and, in (b), we plot the double-occupancy $\left<n_{\uparrow}n_{\downarrow}\right>$ at inverse temperatures $\beta=2$, $4$ and $6$. Simulations were done with a selection of bond dimensions from a range of $14 \le D \le 29$.
\label{fig:benchmarks_eat} }
\end{figure}

\section{Benchmarks}
\label{app:benchmarks}

In order to demonstrate that our algorithm works properly, we collect a series of  benchmark results. The evolution is performed in Trotter steps of size $d\beta = 0.005$, which we found was small enough for step-size independence.  The same time step was used for the results presented in the main text. 

To begin with, we consider non-interacting spinless fermions for which we can compare with analytical results:
\be
H_0 =
- \sum_{\langle i,j \rangle} 
  \left( c_{i}^\dag c_{j} + c_{j}^\dag c_{i} \right).
\label{H0}
\ee
We pushed our simulations up to $\beta=6$.
Fig.~\ref{fig:benchmarks}(a) shows that, as expected, the NTU error~\eqref{NTUerror} during the evolution decreases with increasing total bond dimension. 
Fig.~\ref{fig:benchmarks}(b) shows the relative error of energy as a function of the integrated NTU error~\eqref{sigmadelta}, which in App.~\ref{app:NTU} was argued to be a useful measure of evolution error. Here we define the relative error as $|\frac{E_{\text{iPEPS}} - E_{exact}}{E_{\text{exact}}}|$, where $E_{\text{iPEPS}}$ is the energy from our simulations and $E_{\text{exact}}$ is the exact energy of the Fermi sea. We see a systematic trend where the energy error decreases with decreasing integrated NTU error $\Delta$. It demonstrates the usefulness of $\Delta$ as an error estimator.  

\begin{figure}[t!]
\vspace{-0cm}
\includegraphics[width=\columnwidth,clip=true]{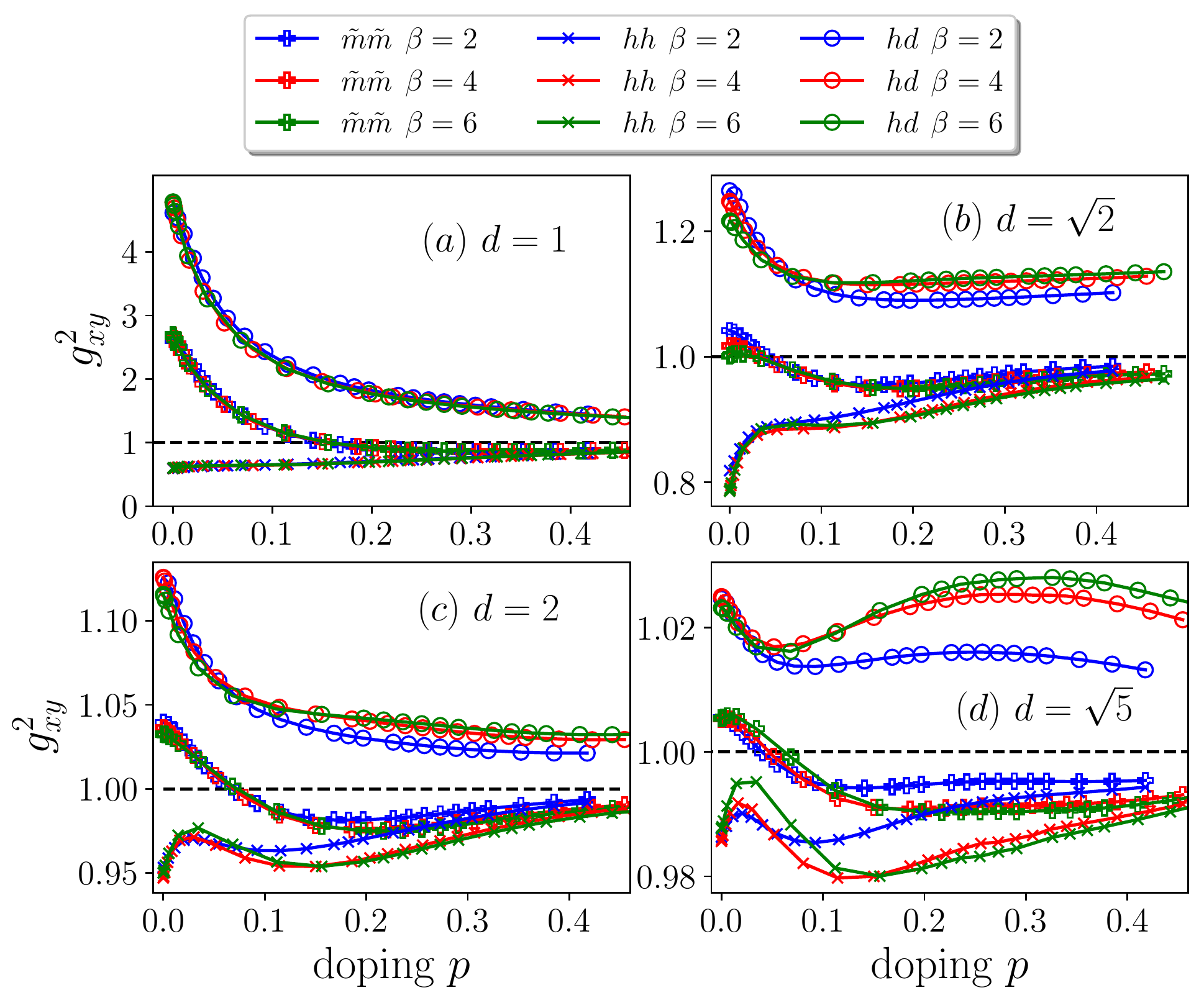}
\vspace{-0cm}
\caption{
{\bf Normalized charge correlators vs doping. } 
We show normalized (a) nearest axial ($d=1$), (b) nearest diagonal ($d=\sqrt{2}$), (c) next nearest axial ($d=2$), and (d) next nearest diagonal ($d=\sqrt{5}$) charge correlators for $\beta=2, 4$ and $6$. Here,   $g^{2}_{\tilde{m}\tilde{m}}$ is the normalized anti-moment correlator, $g^{2}_{hh}$ is the normalized hole-hole correlators and $g^{2}_{hd}$ is the normalized hole-doublon correlator. For detailed definitions, see the main text. We find anti-bunching (bunching) of  $g^{2}_{hd}$ ($g^{2}_{hh}$) for all values of doping $p$ and distances $d$  considered here, although its magnitude decreases with distance. 
Another interesting feature is temperature independence of the correlators for  $d=1$.} 
\label{fig:charge_normalized}
\end{figure}

\begin{figure}[t!]
\vspace{-0cm}
\includegraphics[width=\columnwidth,clip=true]{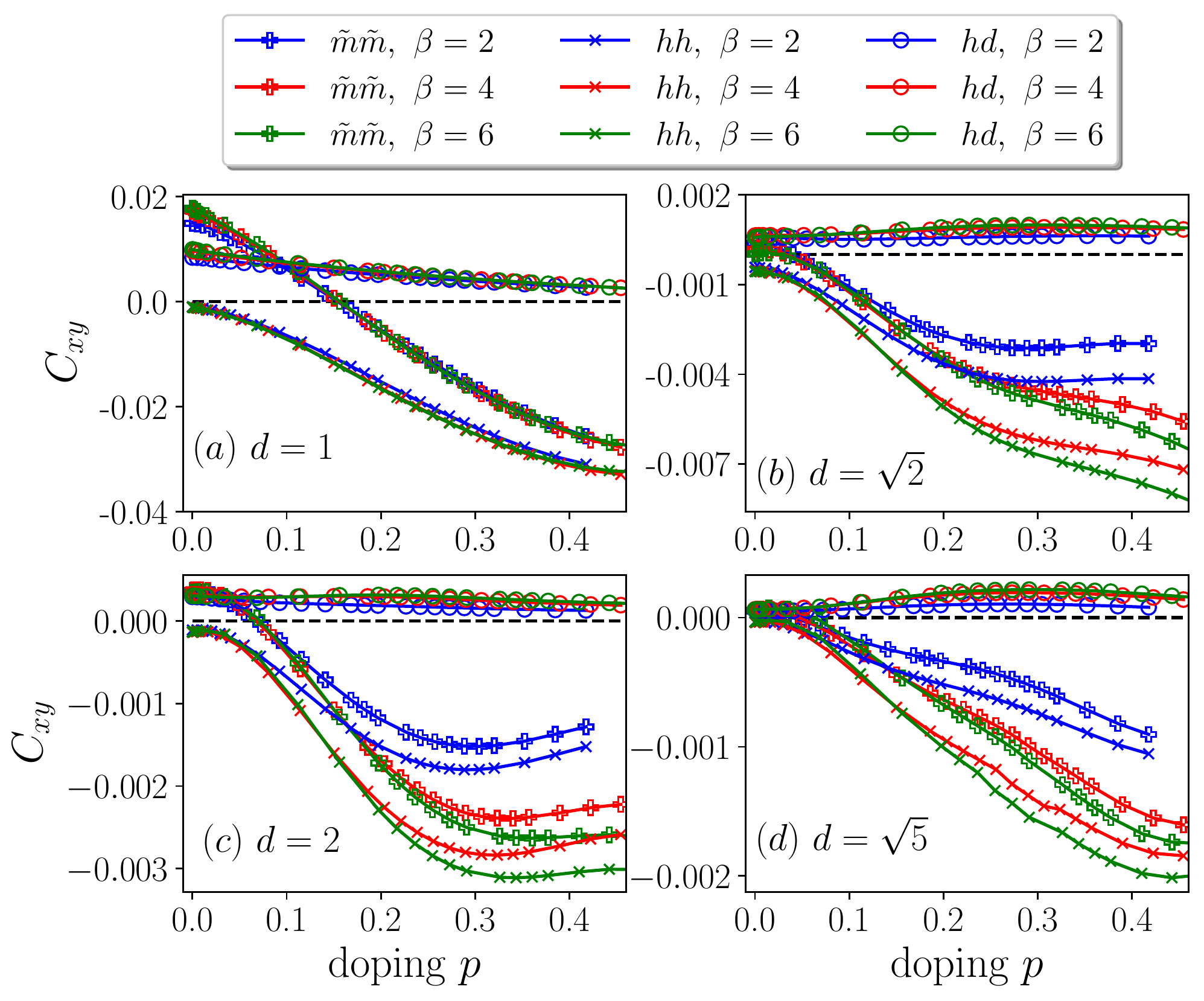}
\vspace{-0cm}
\caption{
{\bf Connected charge correlators vs doping. } 
We show connected two-point charge correlators $C_{\tilde{m}\tilde{m}}$, $C_{hh}$,  $C_{hd}$, for various distances $d$ and inverse temperatures $\beta=2, 4$ and $6$. In  (a) nearest axial ($d=1$), in  (b) nearest diagonal ($d=\sqrt{2}$),  in (c) next nearest axial ($d=2$), and in (d) next nearest diagonal ($d=\sqrt{5}$)  correlators.}
\label{fig:charge_connected}
\end{figure}

Next, we move on to the FHM, where we enforce average dopings of $p=0$ and $p=0.125$ fermion per site and consider the strongly interacting case of $U=8$. In Fig.~\ref{fig:benchmarks}, we plot and compare the energy $E$ per site and double occupancy $\langle n_{\uparrow} n_{\downarrow}\rangle$ for $\beta=2$ and $4$ with the dynamical cluster approximation (DCA) results~\cite{leblanc15}. Our results are in good agreement with DCA. Additionally, we see similar quality of convergence for $\beta=6$ as for $\beta=4$ as a function of decreasing $\Delta$. This boosts confidence in our results for correlators obtained at $\beta=6$ in the main text. Here, we set $\mu=0$ to enforce $p=0$, and to fix $p=0.125$ we scan and fine-tune different chemical potentials, see Tab.~\ref{table:1}. 

Finally, in Fig.~\ref{fig:benchmarks_eat} EAT yields comparable quality of results as FIX in the Hubbard model at half-filling, mutually corroborating both simulation strategies.


\begin{table}[b]
\centering
\begin{tabular}{|| c | c | c | c ||} 
\hline
$D$ & $\beta=2$ & $\beta=4$ & $\beta=6$ \\  
\hline\hline
 14 & -2.2 & -2.247 & -2.277 \\ 
 15 & -2.164 & -2.17 & -2.18 \\
 16 & -2.172 & -2.164 & -2.167 \\
 20 & -2.176 & -2.16 & -2.158 \\
 25 & -2.18 & -2.17 & -2.17 \\ 
 \hline
\end{tabular}
\caption{Chemical potentials used for fixing doping $p=0.125$ for different bond dimensions and inverse temperatures.}
\label{table:1}
\end{table}

\section{Additional data for charge correlators}
\label{app:corr}
For interested readers, we provide additional results for charge correlators. In Fig.~\ref{fig:charge_normalized}, we plot normalized hole-hole $g^2_{hh}$, hole-doublon  $g^2_{hd}$, and anti-moment $g^2_{\tilde{m}\tilde{m}}$  correlators for three values of inverse temperatures $\beta=2, 4$ and $6$ (in the main text, we only provide the data for $\beta=6$ for clarity): 
\begin{equation}
g^2_{xy}(d) = 
\frac{\langle x_i  y_{i+d} \rangle}
{\langle x_i \rangle  \langle y_{i+d} \rangle},
\end{equation}
In Fig.~\ref{fig:charge_connected}, we show connected correlators $C_{xy}$ for the same observables:
\begin{equation}
C_{xy}(d) = \langle x_i  y_{i+d} \rangle - \langle x_i \rangle  \langle y_{i+d} \rangle.
\end{equation} 
Interestingly, the longer the range of the two-point correlator, the stronger the temperature dependence, while for nearest neighbor correlators, $d=1$, there is no discernible dependence on temperature, see Fig.~\ref{fig:charge_normalized}(a) and Fig.~\ref{fig:charge_connected}(a).  
We use iPEPS bond dimension $D=20$ and environmental bond dimension $\chi=120$ for calculation of correlators. The parameters used were found to be sufficient to achieve convergence against bond dimension.

\section{Shifting the particle density}
\label{app:interpolation}

The purification obtained by imaginary time evolution in $\beta$ is---up to errors inflicted by the truncation of bond dimensions after every Trotter gate---equal to $e^{-\frac12\beta H}$. The evolution is performed with a fixed chemical potential $\mu$. The simulation can be repeated for different values of $\mu$, but in general, it is not known beforehand what $\mu$ has to be adopted for a given $\beta$ to reach the desired doping, say, $p=0.125$.  One can bypass this problem by performing evolutions for a grid of $\mu$ and then ``interpolating'' to the $\mu$ that yields the desired particle density. At first sight, the interpolation is rather simple because the total particle number, $N$, commutes with the Hamiltonian. Therefore, knowing $e^{-\frac12\beta H}$ for a given $\mu$, we can obtain a purification for $\mu+\delta\mu$ and the same $\beta$ simply by applying $e^{+\frac12\beta \delta\mu N}$ to the physical indices of the iPEPS. This transformation can be conveniently implemented by applying a local operator $e^{\beta(n_{\uparrow}+n_{\downarrow})\delta\mu/2}$ to the physical index of each purification tensor.

In practice, one has to be cautious because the purification, $e^{-\frac12\beta H}$, is approximated by a tensor network whose bond dimension was truncated after each Trotter gate. The truncation was optimized to minimize the error for a given $\mu$, but the same truncation may turn out not to be optimal for $\mu+\delta\mu$ when $\delta\mu$ gets too large. To be more specific, density operator $\rho$ has a particle number distribution, $f_{\beta,\mu}(N)$. It is reasonable to assume that for given $\beta$ and $\mu$ the truncations were optimized to minimize the error of the dominant central part of the distribution as the optimized cost function had little sensitivity to the errors of its tails, and the relative errors of the tails may remain large. After the transformation we obtain
\begin{equation}
f_{\beta,\mu+\delta\mu}(N)\propto e^{\beta N\delta\mu} f_{\beta,\mu}(N).
\label{eq:den_transf}
\end{equation}
The exponential prefactor shifts the maximum of the new distribution comparing to the old one. When the new maximum is within the error-afflicted tail of the old distribution, the large prefactor magnifies the tail errors. The new distribution fails to be accurate in its new central part, though it remains unreasonably precise in the old central part, which is now an irrelevant tail. This happens when $\delta\mu$ is too large. 

What does it mean too large and how does acceptable $\delta\mu$ depend on $\beta$? We expect that for sufficiently large $\beta$, the distribution localizes on the ground state, which has definite $N$, and $f_{\beta,\mu}(N)$ has very small variance in this regime. Therefore, at sufficiently low temperatures, the allowed magnitude of $\delta\mu$ decreases with increasing $\beta$. The lower are the temperatures at which we want to target a predefined particle density, the finer must be a grid in the chemical potential on which we generate the $\beta$-evolutions. 
\begin{figure}[b]
\vspace{-0cm}
\includegraphics[width=\columnwidth,clip=true]{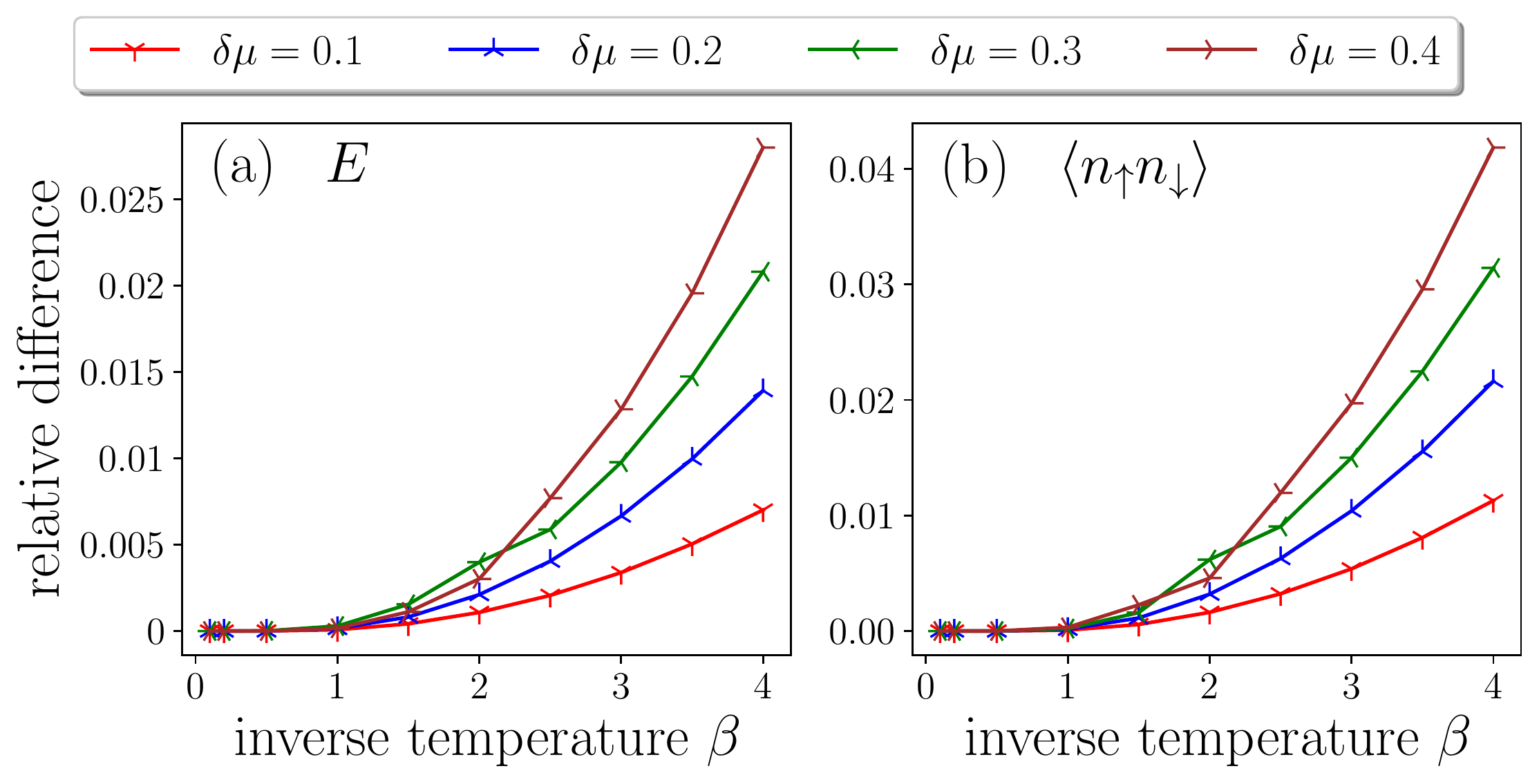}
\vspace{-0cm}
\caption{
{\bf  Error caused by shifting particle density. } 
We test our algorithm for shifting particle density described in App.~\ref{app:interpolation} by plotting the relative difference in expectation values: $|\frac{O^{int}_{\mu+\delta\mu} - O_{\mu}}{O_{\mu}}|$. Here $O_{\mu}$ is the expectation value calculated for a state at a chemical potential $\mu$ and $O^{int}_{\mu+\delta\mu}$ is the expectation value in a state at chemical potential $\mu+\delta\mu$ shifted to $\mu$. The relative difference in observables energy $E$ and double-occupancy $\left<n_{\uparrow}n_{\downarrow}\right>$ are shown in panels (a) and (b), respectively, for $\mu=-2.5$ and $\delta\mu=0.1, 0.2, 0.3, 0.4$. These computations were done for the Hubbard model at $U=8$.}
\label{fig:interp}
\end{figure}

To see if the grid is fine enough, we can make cross-checks between $\mu$ and $\mu+\Delta\mu$, where $\Delta\mu$ is the grid resolution, calculating an observable either directly in the purification at $\mu$ or in the purification at $\mu+\Delta\mu$ transformed back to $\mu$. We corroborate the discussion in Fig.~\ref{fig:interp}, where we plot the relative difference in observable, defined by $|\frac{O^{int}_{\mu+\delta\mu} - O_{\mu}}{O_{\mu}}|$, where $O_{\mu}$ is the expectation value calculated at a chemical potential $\mu$ and $O^{int}_{\mu+\delta\mu}$ is the expectation value shifted from $\mu+\delta\mu$ to $\mu$. Qualitatively the differences depend on $\beta$ and $\delta\mu$ as predicted, adding confidence to the rationale behind the method. For $\delta\mu=0.1$, which is still quite large, the differences are small. 

Since, in our simulations, the NTU evolution was much cheaper than the calculation of expectation values (that employs corner transfer matrix renormalization), we did not use the $\mu$-interpolation. We could afford to generate a fine enough $\mu$-grid to avoid unnecessary interpolation errors.  

\bibliography{ref.bib} 

\end{document}